\documentclass[journal, twocolumn]{IEEEtran}

\usepackage[utf8]{inputenc} 	

\usepackage{amssymb}
\usepackage{bbm} 
\usepackage{graphicx}
\usepackage{color,psfrag}

\usepackage{MnSymbol}

\usepackage[normalem]{ulem}

\usepackage{tikz}
\usepackage{siunitx}
\usepackage{cite} 

\usepackage{longtable}

\usepackage{xcolor}
\newcommand{\tc}[1]{#1}

\newcommand{\e}[2]{{\mathbb E}_{#1}\left[ #2 \right]}
\newcommand{\s}[2]{{\frac{1}{{#1}}\sum_{n=1}^{#1}} {#2}}

\newcommand{\p}{\mathbb P}
\newcommand{\sub}[1]{_{\text{#1}}}
\newcommand{\supe}[1]{^{\text{#1}}}
\newcommand{\pd}{\text{P}\sub{d}}

\newcommand{\pfa}{\text{P}\sub{fa}}

\newcommand{\phz}{\p(\mathcal{H}_0)}
\newcommand{\pho}{\p(\mathcal{H}_1)}

\newcommand{\pdd}{\bar{\text{P}}\sub{d}}

\newcommand{\prcvd}{P\sub{Rx,ST}}
\newcommand{\prcvdsr}{P\sub{Rx,SR}}
\newcommand{\eprcvd}{\hat{P}\sub{Rx,ST}}
\newcommand{\eprcvdsr}{\hat{P}\sub{Rx,SR}}

\newcommand{\ptranst}{P\sub{Tx,ST}}
\newcommand{\ptranpt}{P\sub{Tx,PT}}

\newcommand{\yrcvd}{y\sub{ST}}

\newcommand{\xp}{x\sub{PT}}
\newcommand{\xs}{x\sub{ST}}

\newcommand{\ys}{y\sub{SR}}
\newcommand{\ls}{\lambda\sub{}}
\newcommand{\Ks}{N\sub{s}}

\newcommand{\Kp}{N\sub{p,2}}

\newcommand{\ap}{a\sub{2}}
\newcommand{\bp}{b\sub{2}}
\newcommand{\as}{a\sub{1}}
\newcommand{\bs}{b\sub{1}}

\newcommand{\rs}{R\sub{s}}

\newcommand{\trs}{{R}\sub{s}}
\newcommand{\trsac}{{R}\sub{s}\supe{}}
\newcommand{\trsoc}{{R}\sub{s}\supe{}}

\newcommand{\hpo}{h\sub{p,1}}
\newcommand{\hpt}{h\sub{p,2}}
\newcommand{\hs}{h\sub{s}}
\newcommand{\phpo}{|h\sub{p,1}|^2}
\newcommand{\phpt}{|h\sub{p,2}|^2}
\newcommand{\phs}{|h\sub{s}|^2}
\newcommand{\ehs}{\hat{h}\sub{s}}
\newcommand{\npo}{\sigma^2_{w}}
\newcommand{\spo}{\sigma^2_{x}}
\newcommand{\evar}{\frac{\npo}{2 \Ks}}
\newcommand{\fsam}{f\sub{s}}

\newcommand{\test}{\tau\sub{est}}
\newcommand{\ttest}{\tilde{\tau}\sub{est}}
\newcommand{\tsen}{\tau\sub{sen}}
\newcommand{\ttsen}{\tilde{\tau}\sub{sen}}
\newcommand{\ttsenac}{\tilde{\tau}\sub{sen}\supe{}}
\newcommand{\ttsenoc}{\tilde{\tau}\sub{sen}\supe{}}

\newcommand{\cz}{\text{C}_0}
\newcommand{\co}{\text{C}_1}

\newcommand{\mpd}{\kappa}

\newcommand{\snrrcvd}{{\gamma}\sub{p,1}}
\newcommand{\snrso}{{\gamma}\sub{s}}
\newcommand{\snrpt}{{\gamma}\sub{p,2}}

\newcommand{\lambdas}{\frac{\sigma_w^4}{ 2 \Ks \ptranst}}
\newcommand{\lambdasinv}{\frac{2 \Ks \ptranst}{\sigma_w^4}}

\newcommand{\fpd}{F_{\pd}}
\newcommand{\feprcvd}{F_{\eprcvd}}
\newcommand{\fcz}{F_{\cz}}
\newcommand{\fco}{F_{\co}}

\newcommand{\dsnrs}{f_{\frac{ |\ehs|^2 \ptranst}{\npo}}}
\newcommand{\dsnrp}{f_{\frac{\eprcvdsr}{\npo}}}
\newcommand{\dsnrsp}{f_{\frac{|\ehs|^2 \ptranst}{\npo}\Big /\frac{\eprcvdsr}{\npo}}}
\newcommand{\dcz}{f_{\cz}}
\newcommand{\dco}{f_{\co}}

\newcommand{\thrac}{\mu\supe{}}
\newcommand{\throc}{\mu\supe{}}

\DeclareMathOperator*{\argmaxi}{argmax}
\DeclareMathOperator*{\maxi}{max}

\DeclareMathOperator*{\cchi2}{\mathcal{X}^2}
\DeclareMathOperator*{\ncchi2}{\mathcal{X}_1^2}
\DeclareMathOperator*{\ts}{\text{T}(\textbf{y})}

\newtheorem{theorem}{Theorem}

\newtheorem{lemma}{Lemma}

\newtheorem{remark}{Remark}
\newtheorem{coro}{Corollary}

\makeatletter
\if@twocolumn
	\newcommand{\figscale}{0.9 \columnwidth}
\else
	\newcommand{\figscale}{0.46 \columnwidth}
\fi
\makeatother

\ifCLASSOPTIONcompsoc
\usepackage[caption=false,font=normalsize,labelfont=sf,textfont=sf]{subfig}
\else
\usepackage[caption=false,font=footnotesize]{subfig}
\fi

\usepackage[final=true]{hyperref}

\begin{document}
\title{Sensing-Throughput Tradeoff for Interweave Cognitive Radio System: A Deployment-Centric Viewpoint}
\author{Ankit Kaushik\IEEEauthorrefmark{1}, \IEEEmembership{Student Member, IEEE}, Shree Krishna Sharma\IEEEauthorrefmark{2},  \IEEEmembership{Member, IEEE},\\  Symeon Chatzinotas\IEEEauthorrefmark{2}, \IEEEmembership{Senior Member, IEEE}, Bj\"orn Ottersten\IEEEauthorrefmark{2}, \IEEEmembership{Fellow, IEEE}, \\ Friedrich K. Jondral\IEEEauthorrefmark{1} \IEEEmembership{Senior Member, IEEE} 
\thanks{\IEEEauthorrefmark{1}A. Kaushik and F. K. Jondral are with Communications Engineering Lab, Karlsruhe Institute of Technology (KIT), Germany. Email:{\{ankit.kaushik,friedrich.jondral\}@kit.edu.}} 
\thanks{\IEEEauthorrefmark{2}S.K. Sharma, S. Chatzinotas and B. Ottersten are with SnT - securityandtrust.lu, University of Luxembourg, Luxembourg. Email:{\{shree.sharma, symeon.chatzinotas, bjorn.ottersten\}@uni.lu}.} 
\thanks{The preliminary analysis of this paper has been presented at CROWNCOM 2015 in Doha, Qatar \cite{Kaushik15_CC}.}
\thanks{This work was partially supported by the National Research Fund, Luxembourg under the CORE projects ``SeMIGod'' and ``SATSENT''.} 
}

\maketitle
\thispagestyle{empty}
\pagestyle{empty}

\begin{abstract}
Secondary access to the licensed spectrum is viable only if the interference is avoided at primary system. In this regard, different paradigms have been conceptualized in the existing literature. Among these, interweave systems (ISs) that employ spectrum sensing have been widely investigated. 
Baseline models investigated in the literature characterize the performance of the IS in terms of a sensing-throughput tradeoff, however, this characterization assumes perfect knowledge of involved channels at secondary transmitter, which is unavailable in practice. 
Motivated by this fact, we establish a novel approach that incorporates channel estimation in the system model, 
 and consequently investigate the impact of imperfect channel knowledge on the performance of the IS. More particularly, the variation induced in the detection probability affects the detector's performance at the secondary transmitter, which may result in severe interference at the primary receivers. In view of this, we propose to employ average and outage constraints on the detection probability, in order to capture the performance of the IS.
Our analysis reveals that with an appropriate choice of the estimation time determined by the proposed approach, the degradation in performance of the IS can be effectively controlled, and subsequently the achievable secondary throughput can be significantly enhanced.
\end{abstract}

\section{Introduction}
We are currently in the phase of conceptualizing the requirements of the fifth generation (5G) of mobile wireless systems. One of the major goals is to improve the areal capacity ($\SI{}{bits/s/m^2}$) by a factor of 1000 \cite{Andrews14}. \tc{To this end, an extension to the already allocated spectrum is of paramount importance.} 
Recently, the spectrum beyond \SI{6}{GHz}, which largely entails the millimeter wave is envisaged as a powerful source of spectrum for 5G wireless systems. However, the millimeter wave technology is still in its initial stage and along with complex regulatory requirements in this regime, it has to address several challenges like propagation loss, low efficiency of radio frequency components such as power amplifiers, small size of the antenna and link acquisition \cite{Rapp13}. Therefore, in order to capture a deeper insight of its feasibility in 5G, it is essential to overcome the aforementioned challenges in the near future.

\tc{Besides the spectrum beyond \SI{6}{GHz}, an efficient utilization of the spectrum below \SI{6}{GHz} presents an alternative solution. The use of the spectrum in this regime (below \SI{6}{GHz}) is fragmented and statically allocated, leading to inefficiencies and the shortage in the availability of spectrum for new services.} However, it is possible to overcome this scarcity if we manage to utilize this radio spectrum efficiently. In this perspective, \tc{cognitive radio (CR) is foreseen as one of the potential contenders that addresses the spectrum scarcity problem. Since its origin by Mitola \textit{et al.} in 1999, this notion has evolved at a significant pace, and consequently has acquired certain maturity.} However, from a deployment perspective, this technology is still in its preliminary phase. In this view, it is necessary to make substantial efforts that enable the placement of this concept over a hardware platform.

An access to the licensed spectrum is an outcome to the paradigm employed by the \tc{secondary user (SU)}. Based on the paradigms described in the literature, all CR systems that provide dynamic access to the spectrum mainly fall under three categories, namely, interweave, underlay and overlay systems \cite{Goldsmith09}. In \tc{interweave systems (ISs)}, the SUs render an interference-free access to the licensed spectrum by exploiting spectral holes in different domains such as time, frequency, space and polarization, whereas underlay systems enable an interference-tolerant access under which the SUs are allowed to use the licensed spectrum (e.g. Ultra Wide Band) as long as they respect the interference constraints of the primary receivers (PRs). Besides that, overlay systems consider the participation of higher layers for enabling the spectral coexistence between two or more wireless networks. Due to its ease of deployment, the IS is mostly preferred not only for performing theoretical analysis but also for practical implementation as well. Motivated by these facts, this paper focuses on the performance analysis of the ISs from a deployment perspective. 

\subsection{Motivation and Related Work}
Spectrum sensing is an integral part of ISs. At the secondary transmitter (ST), sensing is necessary for detecting the presence or the absence of a primary user (PU) signal, thereby protecting the PRs against harmful interference. \tc{A sensing mechanism} at the ST can be accomplished by listening to the signal transmitted by the primary transmitter (PT). For detecting a PU signal, several techniques such as energy detection, matched filtering, cyclostationary and feature-based detection exist \cite{Sharma15, Axell12}. Because of its versatility towards unknown PU signals and its low computational complexity, energy detection has been extensively investigated in the literature \cite{Urkowitz, Kostylev02, Alouini03, Herath09, Mariani10}. In this technique, the decision is accomplished by comparing the power received at the ST to a decision threshold. In reality, the ST encounters variations in the received power due to the existence of thermal noise at the receiver and channel fading. Subsequently, these variations lead to sensing errors described as misdetection or false alarm, 
which limit the performance of the IS. In order to determine the performance of a detector, it is essential to obtain the expressions of detection probability and false alarm probability.

In particular, detection probability is critical for ISs because it protects the PR from the interference induced by the ST. As a result, the ISs have to ensure that they operate above a target detection probability \cite{peh07}. Therefore, the characterization of the detection probability becomes absolutely necessary for the performance analysis of the IS. In this context, Urkowitz \cite{Urkowitz} introduced a probabilistic framework for characterizing the sensing errors, however, the characterization accounts only for the noise in the system. To encounter the variation caused by channel fading, a frame structure has been introduced in \cite{Liang08} \tc{assuming that the channel remains} constant over the frame duration, however, upon exceeding the frame duration, the system may observe a different realization of the channel. \tc{Based on this frame structure, the performance of the IS has been investigated in terms of deterministic channel \cite{Liang08, Sharma14, Pradhan15} and random channel\footnote{\tc{In the literature, deterministic and random channels are interpreted as path-loss and fading channels, respectively.}} \cite{Kostylev02, Alouini03, Herath09}. Complementing the analysis in \cite{Liang08, Sharma14, Pradhan15}, in this paper, we consider the involved channels to be deterministic.}

Besides the detection probability, false alarm probability has a large influence on the achievable throughput of the secondary system. 
Recently, the performance characterization of CR systems in terms of a sensing-throughput tradeoff has received significant attention \cite{Liang08, Juarez11, Sharkasi12, Pradhan15}. According to Liang \textit{et al.} \cite{Liang08}, the ST assures a reliable detection of a PU signal by retaining the detection probability above a desired level with an objective of maximizing the throughput at the secondary receiver (SR). In this way, the sensing-throughput tradeoff depicts a suitable sensing time that achieves a maximum secondary throughput. However, to characterize the detection probability and the secondary throughput, the system requires the knowledge of interacting channels, namely, a \textit{sensing} channel, an \textit{access} channel and an \textit{interference} channel, \tc{refer to} \figurename~\ref{fig:scenario}\footnote{As the interference to the PR is controlled by a regulatory constraint over the detection probability, in this view, the interaction with the PR is excluded in the considered scenario \cite{Liang08}.}. To the best of authors' knowledge, the baseline models investigated in the literature assume the knowledge of these channels to be available at the ST. 
However, in practice, this knowledge is not available, thus, needs to be estimated by the secondary system. As a result, from a deployment perspective, the existing solutions for the IS are considered inaccurate for the performance analysis.

\tc{In practice, the knowledge about the involved channels can be estimated either (i) directly by using the conventional channel estimation techniques such as training sequence based \cite{Stoica03} and pilot based \cite{Gifford05, Gifford08} channel estimation or (ii) indirectly by estimating the received signal to noise ratio \cite{Chav11, Sharma13}. It is worthy to note that the sensing and interference channels represent the channels between two different (primary and secondary) systems. In this context, it becomes challenging to select the estimation methods in such a way that low complexity and versatility (towards different PU signals) requirements are satisfied. These issues, discussed later in Section \ref{ssec:pa}, render the existing estimation techniques \cite{Stoica03, Gifford05, Gifford08, Chav11, Sharma13} unsuitable for hardware implementations. 
To this end, we propose to employ a received power based estimation at the ST and at the SR for the sensing and interference channels, respectively.
Considering the fact that the access channel corresponds to the link between the ST and the SR, we propose to employ conventional channel estimation techniques such as pilot based channel estimation at the SR.}

\tc{  
Inherent to the estimation process, the variations due the channel estimation translate to variations in the performance parameters, namely detection probability and secondary throughput. In particular, the variations induced in the detection probability may result in harmful interference at the PR, hence, severely degrading the performance of a CR system. In this context, the performance characterization of an IS with imperfect channel knowledge remains an open problem. 
In this regard, this paper focuses on the performance characterization of the IS in terms of sensing-throughput tradeoff taking these aforementioned aspects into account.}

\subsection{Contributions}
The major contributions of this paper can be summarized as follows:
\subsubsection{\tc{Analytical framework}}
\tc{In contrast to the existing models that assume the perfect knowledge of the channels, the main goal of this paper is to derive an analytical framework that constitutes the estimation of: (i) sensing channel at the ST, (ii) access channel and (iii) interference channel at the SR. Under this framework, we propose a novel integration of the channel estimation in the secondary system's frame structure, according to which, we take into account the samples considered for channel estimation (of the sensing channel) also for sensing in such a way that the time resources within the frame are utilized efficiently. Furthermore, we select the estimation techniques in such a way that the hardware complexity and the versatility towards unknown PU signals requirements (as considered while employing an energy based detection) are not compromised. In this context, we propose to employ a received power based estimation for the sensing and interference channels. Based on this framework, we characterize the performance of the IS by considering: (i) the variations due to imperfect channel knowledge and (ii) the performance degradation due to the inclusion of channel estimation.}

\subsubsection{\tc{Imperfect channel knowledge}}
\tc{To capture the variations induced due to imperfect channel knowledge, we characterize the distribution functions of performance parameters such as detection probability and achievable secondary throughput. More importantly, we utilize the distribution function of the detection probability to incorporate two primary user (PU) constraints, namely, average and outage constraints on the detection probability. In this way, the proposed approach is able to control the amount of excessive interference caused at the PR due to the imperfect channel knowledge.} 
\subsubsection{\tc{Estimation-sensing-throughput tradeoff}}
Subject to the average and the outage constraints, we establish the expressions of sensing-throughput tradeoff that capture the aforementioned variations and evaluate the performance loss in terms of the achievable secondary throughput. \tc{In particular, we propose two different optimization approaches for countering the variations in sensing-throughput tradeoff and determining a suitable sensing time, which attains a maximum secondary throughput.} Finally, we depict a fundamental tradeoff between estimation time, sensing time and achievable secondary throughput. We exploit this tradeoff to determine a suitable estimation and sensing time that depicts the maximum achievable performance of the IS. 

\subsection{Organization}
The subsequent sections of the paper are organized as follows: Section \ref{sec:sys_mod} describes the system model that includes the deployment scenario and the signal model. \tc{Section \ref{sec:pr_mod} presents the problem description and the proposed approach.} Section \ref{sec:ana} characterizes the distribution functions of the performance parameters and establishes the sensing-throughput tradeoff subject to average and outage constraints. Section \ref{sec:num_ana} analyzes the numerical results based on the obtained expressions. Finally, Section \ref{sec:conc} concludes the paper. Table \ref{tb:tb1} lists the definitions of acronyms and important mathematical notations used throughput the paper.

\begin{table}
\renewcommand{\arraystretch}{1.4}
\caption{Definitions of Acronyms and Notations used}
\label{tb:tb1}
\centering
\scriptsize{
\begin{tabular}{p{0.25\columnwidth}||p{0.6\columnwidth}}
\hline
\bfseries Acronyms and Notations & \bfseries Definitions \\
\hline\hline
AC, OC & average constraint, outage constraint \\ \hline
CR & cognitive radio\\ \hline
CSC, CSC-BS, MC-BS, MS & cognitive small cell, cognitive small cell-base station, macro cell-base station, mobile station\\ \hline
IM, EM & ideal model, estimation model \\ \hline
IS & interweave system \\ \hline
PU - PT, PR & primary user - primary transmitter, primary receiver \\ \hline
SU - ST, SR & secondary user - secondary transmitter, secondary receiver \\ \hline
$\mathcal H_1, \mathcal H_0$ & Signal plus noise hypothesis, noise only hypothesis\\ \hline
$\fsam$ & Sampling frequency\\ \hline
$\test, \tsen$ & Estimation time, sensing time interval\\ \hline
$T$ & Frame duration\\ \hline
$\pd, \pfa$ & Detection probability, false alarm probability \\ \hline
$\pdd$ & Target detection probability\\ \hline
$\mpd$ & Outage constraint over detection probability\\ \hline
$\hpo, \hpt, \hs$ & Channel coefficient for the link PT-ST, PT-SR, ST-SR \\ \hline
$\snrrcvd, \snrso$ & Signal to noise ratio for the link PT-ST, ST-SR \\ \hline
$\snrpt$ & Interference (from PT) to noise ratio for the link PT-SR \\ \hline
$\rs$ & Throughput at SR\\ \hline
$\cz,\co$ & Date rate at SR without and with interference from PT  \\ \hline
$\mu$ & Threshold for the energy detector\\ \hline
$F_{(\cdot)}$ & Cumulative distribution function of random variable $(\cdot)$\\ \hline
$f_{(\cdot)}$ & Probability density function of random variable $(\cdot)$\\ \hline
$\hat{(\cdot)}$ & Estimated value of ($\cdot$)\\ \hline
$\tilde{(\cdot)}$ & Suitable value of the parameter ($\cdot$) that achieves maximum performance \\ \hline
$\mathbb E_{(\cdot)}$ & Expectation with respect to ($\cdot$) \\ \hline
$\p$ & Probability measure \\ \hline
\textbf{T}$(\cdot)$ & Test statistics\\ \hline
$\spo,  \npo$ & Signal variance at PT, noise variance at ST and SR\\ \hline
$ \Ks$ & Number of pilot symbols used for pilot based estimation at the SR for $\hs$ \\ \hline
$ \Kp$ & Number of samples used for received power based estimation at the SR for $\hpt$ \\ \hline
\end{tabular}}
\end{table}
 
\section{System Model} \label{sec:sys_mod}
\subsection{Deployment Scenario}

\begin{figure}[!t]
\centering
\includegraphics[width = \figscale]{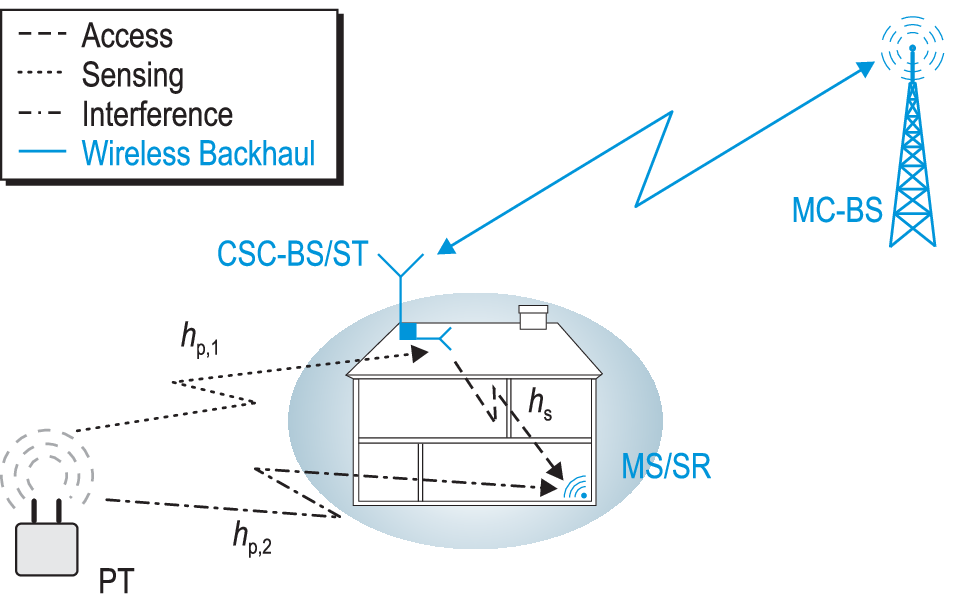}
\caption{A cognitive small cell scenario demonstrating: (i) the interweave paradigm, (ii) the associated network elements, which constitute cognitive small cell-base station/secondary transmitter (CSC-BS/ST), mobile station/secondary receiver (MS/SR), macro cell-base station (MC-BS) and primary transmitter (PT), (iii) the interacting channels: sensing ($\hpo$), access ($\hs$) and interference ($\hpt$).} 
\label{fig:scenario}
\vspace{-0.3cm}
\end{figure}
The cognitive small cell (CSC), a CR application, characterizes a small cell deployment that fulfills the spectral requirements for mobile stations (MSs) operating indoor, \tc{refer to} \figurename~\ref{fig:scenario}. For the disposition of the CSC in the network, the following key elements are essential: a CSC-base station (CSC-BS), a macro cell-base station (MC-BS) and MS, refer to \figurename~\ref{fig:scenario}. MSs are the indoor devices served by the CSC-BS over an access channel ($\hs$). Furthermore, the MC-BS is connected to several CSC-BSs over a wireless backhaul\footnote{A wireless backhaul is a point-to-point wireless link between the CSC-BS and MC-BS that relays the traffic generated from the CSC to the core network.}. Moreover, the transmissions from the PT can be listened by the CSC-BS and the MS over sensing ($\hpo$) and interference channel ($\hpt$), respectively. 
Considering the fact that the IS is employed at the CSC-BS, the CSC-BS and the MS represent ST and SR, respectively. A hardware prototype of the CSC-BS operating as IS was presented in \cite{Kaushik13}. For simplification, a PU constraint based on false alarm probability was considered in \cite{Kaushik13}. With the purpose of improving system's reliability, we extend the analysis to employ a PU constraint on the detection probability. 

\tc{Complementing the analysis depicted in \cite{Liang08},} we consider a slotted medium access for the IS, where the time axis is segmented into frames of length $T$, according to which, the ST employs periodic sensing. Hence, each frame consists of a sensing slot $\tsen$ and the remaining duration $T - \tsen$ is utilized for data transmission. For small $T$ relative to the PUs' expected ON/OFF period, the requirement of the ST to be in alignment to PUs' medium access can be relaxed \cite{Wang09, Tang11, Zhao12}.  
 
\subsection{Signal model}
Subject to the underlying hypothesis that illustrates the presence $(\mathcal{H}_1)$ or absence ($\mathcal{H}_0$) of a PU signal, the discrete and real signal received at the ST is given by  
\begin{equation}
\yrcvd[n] = 
\begin{cases}
\hpo \cdot \xp[n] + w[n] & : \mathcal{H}_1 \\
w[n] & :\mathcal{H}_0
\end{cases},
\label{eq:sys_mod_p1s}
\end{equation}
where $\xp[n]$ corresponds to a discrete and real sample transmitted by the PT, $|\hpo|^2$ represents the power gain of the sensing channel for a given frame and $w[n]$ is additive white Gaussian noise at the ST. \tc{According to \cite{Liang08}, the signal $\xp[n]$ transmitted by the PUs} can be modelled as: (i) phase shift keying modulated signal, or (ii) Gaussian signal. The signals that are prone to high inter-symbol interference or entail precoding can be modelled as Gaussian signals. For this paper, we focus our analysis on the latter case. As a result, the mean and the variance for the signal and the noise are determined as $\e{}{\xp[n]} = 0$, $\e{}{w[n]} = 0$, $\e{}{|\xp[n]|^2} = \spo$ and $\e{}{|w[n]|^2} = \npo$. The channel $\hpo$ is considered to be independent of $\xp[n]$ and $w[n]$, thus, $\yrcvd$ is also an independent and identically distributed (i.i.d.) random process. 

Similar to (\ref{eq:sys_mod_p1s}), during data transmission, the discrete and real received signal at the SR conditioned on the detection probability ($\pd$) and false alarm probability ($\pfa$) is given by
\begin{equation}
\ys[n] = 
\begin{cases}
\hs \cdot \xs[n] + \hpt \cdot \xp[n] + w[n] & : 1 - \pd \\
\hs \cdot \xs[n] + w[n] & : 1 - \pfa
\end{cases},
\label{eq:sys_mod_ss}
\end{equation}
where $\xs[n]$ corresponds to discrete and real sample transmitted by the ST. Further, $|\hs|^2$ and $|\hpt|^2$ represent the power gains for the access and the interference channels, \tc{refer to} \figurename~\ref{fig:scenario}. 

\section{\tc{Problem Description and Proposed Approach}} \label{sec:pr_mod}
\subsection{\tc{Problem Description}}\label{ssec:pd}
\tc{In accordance with the conventional frame structure}, the ST performs sensing for a duration of $\tsen$. The test statistics $\ts$ at the ST is evaluated as   
\begin{align}
\ts = \s{\tsen \fsam}{ |\yrcvd[n]|^2} \mathop{\gtrless}_{\mathcal{H}_0}^{\mathcal{H}_1} \mu, 
\label{eq:test_st}
\end{align}
where $\mu$ is the decision threshold and $\textbf{y}$ is a vector with $\tsen \fsam$ samples. $\ts$ represents a random variable, whereby the characterization of the distribution function depends on the underlying hypothesis. Corresponding to $\mathcal{H}_0$ and $\mathcal{H}_1$, $\ts$ follows a central chi-squared ($\cchi2$) distribution \cite{Kay}. 
As a result, the detection probability $(\pd)$ and the false alarm probability $(\pfa)$ corresponding to (\ref{eq:test_st}) are determined as \cite{Tan08}
\begin{align}
\pd(\mu, \tsen, \prcvd) &= \Gamma\left( \frac{\tsen \fsam}{2}, \frac{\tsen \fsam \mu}{2 \prcvd} \right),  \label{eq:pd} 
\end{align}
\begin{align}
\pfa(\mu, \tsen) &= \Gamma\left( \frac{\tsen \fsam}{2}, \frac{\tsen \fsam \mu}{2 \npo} \right),  \label{eq:pfa} 
\end{align}
where $\prcvd$ is the power received over the sensing channel and $\Gamma(\cdot, \cdot)$ represents a regularized incomplete upper Gamma function \cite{grad}.


Following the characterization of $\pfa$ and $\pd$, Liang \textit{et al.} \cite{Liang08} established a tradeoff between the sensing time and secondary throughput $(\rs)$ subject to a target detection probability $(\pdd)$. This tradeoff is represented as  
\begin{align}
\trs(\ttsen) &= \maxi_{\tsen} \rs(\tsen) = \frac{T- \tsen}{T} \bigg[ \cz (1 - \pfa) \phz + \nonumber \\ \quad & \co (1 - \pd) \pho  \bigg], \label{eq:thr_id} \\
\text{s.t.} & \text{ } \pd \ge \pdd, \label{eq:thr_id_con} \\ 
\text{where } & \cz = \log_2 \left(1 + |\hs|^2 \frac{\ptranst}{\npo}\right) = \log_2 \left( 1 + \snrso \right) \label{eq:Cap0}\\ 
{\text{ and }} & \co = \log_2 \left(1 + \frac{|\hs|^2 \ptranst }{|\hpt|^2 \ptranpt  + \npo} \right) \nonumber \\ \quad & = \log_2 \left(1 + \frac{|\hs|^2 \ptranst }{\prcvdsr} \right) = \log_2 \left(1 + \frac{\snrso}{\snrpt + 1}  \right), \label{eq:Cap1} 
\end{align}
where $\phz$ and $\pho$ are the occurrence probabilities for the respective hypothesis, whereas $\snrpt$ and $\snrso$ correspond to interference (from the PT) to noise ratio and signal to noise ratio for the links PT-SR and ST-SR, respectively. Moreover, $\ptranst$ and $\ptranpt$ represent the transmit power at the PT and the ST, whereas $\prcvdsr$ corresponds to the received power (which includes interference power from the PT and the noise power) at the SR. In addition, $\cz$ and $\co$ represent the data rate without and with interference from the PT. In other words, using (\ref{eq:thr_id}), the ST determines a suitable sensing time $\tsen = \ttsen$, such that the secondary throughput is maximized subject to a target detection probability, \tc{refer to} (\ref{eq:thr_id_con}). From the deployment perspective, the tradeoff depicted above has the following fundamental issues:
\begin{itemize}
\item Without the knowledge of the received power $\prcvd$ over the sensing channel, it is not feasible to characterize $\pd$, refer to (\ref{eq:pd}). This leaves the characterization of the throughput (\ref{eq:thr_id}) impossible and the constraint defined in (\ref{eq:thr_id_con}) inappropriate. 
\item Moreover, the knowledge of the interference and the access channels is required at the ST, \tc{refer to} (\ref{eq:Cap0}) and (\ref{eq:Cap1}) for characterizing the throughput in terms of $\cz$ and $\co$ at the SR. 
\end{itemize}
Taking these issues into account, it is not feasible to employ the performance analysis depicted by this model \tc{(referred as ideal model, hereafter)} for hardware implementation. In the subsequent section, we propose an analytical framework \tc{(also referred as estimation model)} that addresses the aforementioned issues, thereby including the estimation of the sensing channel at the ST, and the interference and the access channels at the SR. Based on the proposed approach, we then investigate the performance of the IS in terms of the sensing-throughput tradeoff. 

\subsection{\tc{Proposed Approach}} \label{ssec:pa}

\tc{In order to overcome the difficulties discussed in Section \ref{ssec:pd}, the following strategy is proposed in this paper. \begin{enumerate}
\item As a first step, we consider the estimation of the involved channels. In order to characterize the detection probability, we propose to employ a received power based estimation at the ST for the sensing channel. This is done to ensure that detection probability remains above a desired level. We further to employ a pilot based estimation and a received power based estimation for the access channel and the interference channel, respectively, at the SR, to characterize the secondary throughput. 
\item Next, we characterize the variations due to channel estimation in the estimated parameters, namely, received power (for the sensing and the interference channels) and the power gain (for the access channel) in terms of their cumulative distribution functions. 
\item In order to investigate the performance of the IS subject to the channel estimation, we further characterize these variations in the performance parameters, which include detection probability and secondary throughput, in terms of their cumulative distribution functions. 
\item Finally, we utilize the derived cumulative distribution functions to obtain the expressions of sensing-throughput tradeoff. Hence, based on these expressions, we quantify the impact of imperfect channel knowledge on performance of the ISs, and subsequently determine the achievable secondary throughput at a suitable sensing time. 
\end{enumerate}
}

It is well-known that systems with transmitter information (which includes the filter parameters, pilot symbols, modulation type and time-frequency synchronization) at the receiver acquire channel knowledge by listening to the pilot data sent by the ST \cite{Gans71, Gifford05, Gifford08, Anna05}. Other systems, where the receiver possesses either no access to this information or limited by hardware complexity, procure channel knowledge indirectly by estimating a different parameter that entails the channel knowledge, for instance, received signal power \cite{Kaushik15_CC} or received signal to noise ratio \cite{Chav11, Sharma13}. Recently, estimation techniques such as pilot based estimation \cite{Suraweera10, Kim12} and received power based estimation \cite{Kaushik15} have been applied to obtain channel knowledge for CR systems. However, the performance analysis has been limited to underlay systems, where the emphasis has been given on modelling the interference at the PR. 

Since the pilot based estimation requires the knowledge of the PU signal at the secondary system, the versatility (in terms of PU signals) of the secondary system is compromised. On the other side, for the estimation of the received signal to noise ratio, Eigenvalue (which involves matrix operations) based approach \cite{Sharma13} or iterative approaches such as expectation-maximization have been proposed \cite{Chav11}. Due to the complicated mathematical operations or the complexity of the iterative algorithms, such approaches tend to increase the hardware complexity of the ISs. In order to resolve these issues, we propose to employ received power based estimation for the sensing and interference channels, and pilot based estimation for the access channel. Similar to the energy based detection, since the received power based estimation involves simple operations on the obtained samples such as magnitude squared followed by summation, the proposed estimation provides a reasonable tradoff between complexity and versatility. 

However, with the inclusion of this estimation, the system anticipates: (i) a performance loss in terms of temporal resources used and (ii) variations in the aforementioned performance parameters due to estimation. A preliminary analysis of this performance loss was carried out in \cite{Kaushik15_CC}, where it was revealed that in low signal to noise ratio regime, imperfect knowledge of received power corresponds to large variation in detection probability, hence, causes a severe degradation in the performance of the IS. However, this performance degradation was determined by means of lower and upper bounds. In this work, we consider a more exact analysis, whereby we capture the variations in detection probability by characterizing its distribution function, and subsequently apply new probabilistic constraints on the detection probability, which allow ISs to operate at low signal to noise ratio regime. 

\begin{figure}[!t]
\centering
\includegraphics[width = \figscale]{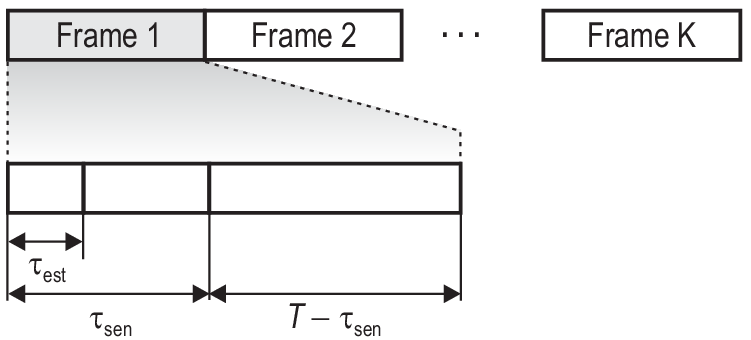}
\caption{\tc{An illustration of the proposed frame structure for an interweave system depicting the estimation phase and the sensing phase for the sensing channel.}}
\label{fig:fs}
\vspace{-0.5cm}
\end{figure}

In order to include channel estimation, we propose a frame structure that constitutes an estimation $\test$, a sensing $\tsen$ and data transmission $T - \tsen$, where $\test$ and $\tsen$ correspond to time intervals and \tc{$0 < \test \le \tsen < T$, refer to} \figurename~{\ref{fig:fs}}. Since the estimated values of the interacting channels are required for determining the suitable sensing time (the duration of the sensing phase), the sequence depicted in \figurename~{\ref{fig:fs}}, whereby estimation followed by sensing is reasonable for the hardware deployment. 
\tc{Particularly for the sensing channel, it is worthy to note that the samples used for estimation can be combined with the samples acquired for sensing\footnote{\tc{Therefore, the sensing phase incorporates the estimation phase, see \figurename~\ref{fig:fs}}.} such that the time resources within the frame duration can be utilized efficiently, as shown in the frame structure in \figurename~\ref{fig:fs}.} 
To avail the estimates for the interference and access channels at the ST, a low-rate feedback channel from the SR to the ST is required for the proposed approach. 
In the following paragraphs, we consider the estimation of the involved channels. 
\subsubsection{Estimation of sensing channel ($\hpo$)}
Following the previous discussions, the ST acquires the knowledge of $\hpo$ by estimating its received power. The estimated received power is required for the characterization of $\pd$, thereby evaluating the detector performance. 

Under $\mathcal H_1$, the received power based estimated during the estimation phase at the ST is given as \cite{Urkowitz} 
\begin{align}
\eprcvd = \s{\test \fsam}{ |\yrcvd[n]|^2}.
\label{eq:eprcvd} 
\end{align}
$\eprcvd$ determined in (\ref{eq:eprcvd}) using $\test \fsam$ samples follows a central chi-squared distribution $\cchi2$ \cite{Kay}. 
The cumulative distribution function (CDF) of $\eprcvd$ is given by  
\begin{align}
\feprcvd(x) = 1 - \Gamma\left(\frac{\test \fsam}{2}, \frac{ \test \fsam x}{2 \prcvd}  \right). 
\label{eq:dprcvd}
\end{align}


\subsubsection{Estimation of access channel ($\hs$)}
The signal received from the ST undergoes matched filtering and demodulation at the SR, hence, it is reasonable to employ pilot based estimation for $\hs$. Unlike received power based estimation, pilot based estimation renders a direct estimation of the channel. 
Now, to accomplish pilot based estimation, the SR aligns itself to pilot symbols transmitted by the ST. Under $\mathcal H_0$, the discrete and real pilot symbols at the output of the demodulator is given by \cite{Gifford08} 
\begin{align}
p[n] = \sqrt{E\sub{s}} \hs + w[n], 
\label{eq:pilot_sig}
\end{align}
where $E\sub{s}$ denotes the pilot energy. Without loss of generality, the pilot symbols are considered to be +1. The maximum likelihood estimate, representing a sample average of $\Ks$ pilot symbols, is given by \cite{Gifford05}
\begin{align}
\hs = \ehs + \smash[b]{\underbrace{\frac{\sum^{\Ks}_{n} p[n]}{2 \Ks}}_{\epsilon}},
\label{eq:pilot_MLE}
\end{align}\\[-0.3em]
where $\epsilon$ denotes the estimation error. 
The estimate $\ehs$ is unbiased, efficient and achieves a Cram\'er-Rao bound with equality, with variance $\e{}{|\hs -\ehs|^2} = \npo/(2 \Ks)$ \cite{Gifford08}. Consequently, $\ehs$ conditioned on $\hs$ follows a Gaussian distribution.
\begin{align}
\ehs|\hs \sim \mathcal{N}\left( \hs,\evar \right).
\label{eq:ehs} 
\end{align}
As a result, the power gain $|\ehs|^2$ follows a non-central chi-squared ($\ncchi2$) distribution with 1 degree of freedom and non-centrality parameter $\ls = \frac{2 \Ks |\hs|^2}{\npo}$.  

\subsubsection{Estimation of interference channel ($\hpt$)}
Analog to sensing channel, the SR performs received power based estimation by listening to the transmission from the PT. The knowledge of $\hpt$ is required to characterize interference from the PT. Under $\mathcal H_1$, the discrete signal model at the SR is given as 
\begin{align}
\ys[n] = \hpt \cdot \xp[n] + w[n].
\label{eq:sys_mod_p2s}
\end{align}
The received power at the SR from the PT given by  
\begin{align} 
\eprcvdsr &= \frac{1}{\Kp} \sum\limits_{n = 1}^{\Kp} |\ys[n]|^2,
\label{eq:ehp2}
\end{align}
follows a $\cchi2$ distribution, where $\Kp$ corresponds to the number of samples used for estimation.

\subsection{\tc{Validation}}
\tc{
It is now clear that the estimates $\eprcvd$, $|\ehs|^2$ and $\eprcvdsr$ exhibit the knowledge corresponding to the involved channels, however, it is essential to validate them, mainly $\eprcvd$ and $\eprcvdsr$. In this context, it is necessary to ensure the presence of the PU signal ($\mathcal H_1$) for that particular frame. In this direction, Chavali \textit{et al.} \cite{Chav11} recently proposed a detection followed by the estimation of the signal to noise ratio, while \cite{Cao14} implemented a blind technique for estimating signal power of non-coherent PU signals. In this paper, we propose a different methodology, according to which, we apply a coarse detection\footnote{\tc{For the coarse detection, an energy detection is employed whose threshold can be determined by means of a constant false alarm rate.}} on the estimates $\eprcvd$, $\eprcvdsr$ at the end of the estimation phase $\test$. Through an appropriate selection of the time interval $\test$ (for instance, $\test \in [1, 10]\SI{}{ms}$) during the system design, the reliability of the coarse detection can be ensured. With the existence of a separate control channel such as cognitive pilot channel, the reliability of the coarse detection can be further enhanced by exchanging the detection results between the ST and the SR.} 

\tc{
Since the estimation and the coarse detection processes in our proposed method are equivalent in terms of their mathematical operations (which include magnitude squared and summation), we consider the validity of the channel estimates with certain reliability and without comprising the complexity of the estimators employed by the secondary system. Moreover, by performing a joint estimation and (coarse) detection, we propose an efficient way of utilizing the time resources within the frame duration. The ST considers these estimates to determine a suitable sensing time based on the sensing-throughput tradeoff such that the desired detector's performance is ensured. At the end of the detection phase, we carry out fine detection\footnote{\tc{In accordance with the proposed frame structure in Fig. 2, fine detection represents the main detection which also includes the samples acquired during the estimation phase.}} of the PU signals, thereby improving the performance of the detector.} 
  
\subsection{Assumptions and Approximations}
To simplify the analysis and sustain analytical tractability for the proposed approach, several assumptions considered in the paper are summarized as follows:
\begin{itemize}
\item We consider that all transmitted signals are subjected to distance dependent path loss and small scale fading gain. 
With no loss of generality, we consider that the channel gains include distance dependent path loss and small scale gain. Moreover, the coherence time for the channel gain is considered to be greater than the frame duration\footnote{In the scenarios where the coherence time exceeds the frame duration, in such cases, our characterization depicts a lower performance bound.}. 
\item We assume the perfect knowledge of the noise power in the system, however, the uncertainty in noise power can be captured as a bounded interval \cite{Tan08}. Inserting this interval in the derived expressions, \tc{refer to} Section \ref{sec:ana}, the performance of the IS can be expressed in terms of the upper and the lower bounds. 
\item For all degrees of freedom, $\ncchi2$ distribution can be approximated by Gamma distribution \cite{abramo}. The parameters of the Gamma distribution are obtained by matching the first two central moments to those of $\ncchi2$. 
\end{itemize}

\section{Theoretical Analysis} \label{sec:ana}
At this stage, it is evident that the variation due to imperfect channel knowledge translates to the variations of the performance parameters $\pd, \cz$ and $\co$, which are fundamental to sensing-throughput tradeoff. Below, we capture these variations by characterizing their cumulative distribution functions $\fpd$, $\fcz$ and $\fco$, respectively. 
\begin{lemma} \label{lm:lem1}
\normalfont
The cumulative distribution function of $\pd$ is characterized as 
\begin{align}
\fpd(x) = 1 - \Gamma \left(\frac{\test \fsam}{2}, \frac{\test \fsam \tsen \fsam \mu}{4 \prcvd \Gamma^{-1}(x, \frac{\tsen \fsam}{2}) } \right), 
\label{eq:fpd}
\end{align}
where $\Gamma^{-1}(\cdot, \cdot)$ is inverse function of regularized upper Gamma function \cite{grad}.  
\end{lemma} 
\begin{IEEEproof}
The cumulative distribution function of $\pd$ is defined as 
\begin{align}
\fpd(x) = \p(\pd(\mu, \tsen, \eprcvd) \le x).
\end{align}
Using (\ref{eq:pd})
\begin{align}
\quad =  \p \left( \Gamma \left( \frac{\tsen \fsam}{2}, \frac{\tsen \fsam \mu}{2 \eprcvd} \right) \le x \right), 
\end{align}
\begin{align}
\quad =  1- \p \left( \eprcvd \ge \frac{\mu \tsen \fsam}{2 {\Gamma}^{-1}\left( x, \frac{\tsen \fsam}{2} \right) } \right). \label{eq:lem1} 
\end{align}
Replacing the cumulative distribution function of $\eprcvd$ in (\ref{eq:lem1}), we obtain an expression of $\fpd$.
\end{IEEEproof}

\captionsetup[subfigure]{position=top}
\begin{figure}[!t]
\centering
\subfloat[]{
%
%
%
\psfrag{s05}[t][t]{\fontsize{8}{12}\fontseries{m}\mathversion{normal}\fontshape{n}\selectfont \color[rgb]{0,0,0}\setlength{\tabcolsep}{0pt}\begin{tabular}{c}$\pd$\end{tabular}}%
\psfrag{s06}[b][b]{\fontsize{8}{12}\fontseries{m}\mathversion{normal}\fontshape{n}\selectfont \color[rgb]{0,0,0}\setlength{\tabcolsep}{0pt}\begin{tabular}{c}CDF\end{tabular}}%
\psfrag{s10}[][]{\fontsize{10}{15}\fontseries{m}\mathversion{normal}\fontshape{n}\selectfont \color[rgb]{0,0,0}\setlength{\tabcolsep}{0pt}\begin{tabular}{c} \end{tabular}}%
\psfrag{s11}[][]{\fontsize{10}{15}\fontseries{m}\mathversion{normal}\fontshape{n}\selectfont \color[rgb]{0,0,0}\setlength{\tabcolsep}{0pt}\begin{tabular}{c} \end{tabular}}%
\psfrag{s12}[l][l]{\fontsize{8}{12}\fontseries{m}\mathversion{normal}\fontshape{n}\selectfont \color[rgb]{0,0,0}Simulated}%
\psfrag{s13}[l][l]{\fontsize{8}{12}\fontseries{m}\mathversion{normal}\fontshape{n}\selectfont \color[rgb]{0,0,0}Theoretical}%
\psfrag{s14}[l][l]{\fontsize{8}{12}\fontseries{m}\mathversion{normal}\fontshape{n}\selectfont \color[rgb]{0,0,0}Simulated}%
%
\fontsize{8}{12}\fontseries{m}\mathversion{normal}%
\fontshape{n}\selectfont%
%
\psfrag{x01}[t][t]{0}%
\psfrag{x02}[t][t]{0.1}%
\psfrag{x03}[t][t]{0.2}%
\psfrag{x04}[t][t]{0.3}%
\psfrag{x05}[t][t]{0.4}%
\psfrag{x06}[t][t]{0.5}%
\psfrag{x07}[t][t]{0.6}%
\psfrag{x08}[t][t]{0.7}%
\psfrag{x09}[t][t]{0.8}%
\psfrag{x10}[t][t]{0.9}%
\psfrag{x11}[t][t]{1}%
%
\psfrag{v01}[r][r]{0}%
\psfrag{v02}[r][r]{0.1}%
\psfrag{v03}[r][r]{0.2}%
\psfrag{v04}[r][r]{0.3}%
\psfrag{v05}[r][r]{0.4}%
\psfrag{v06}[r][r]{0.5}%
\psfrag{v07}[r][r]{0.6}%
\psfrag{v08}[r][r]{0.7}%
\psfrag{v09}[r][r]{0.8}%
\psfrag{v10}[r][r]{0.9}%
\psfrag{v11}[r][r]{1}%
%
%

\begin{tikzpicture}[scale=1]
\node[anchor=south west,inner sep=0] (image) at (0,0)
{
	\includegraphics[width = \figscale]{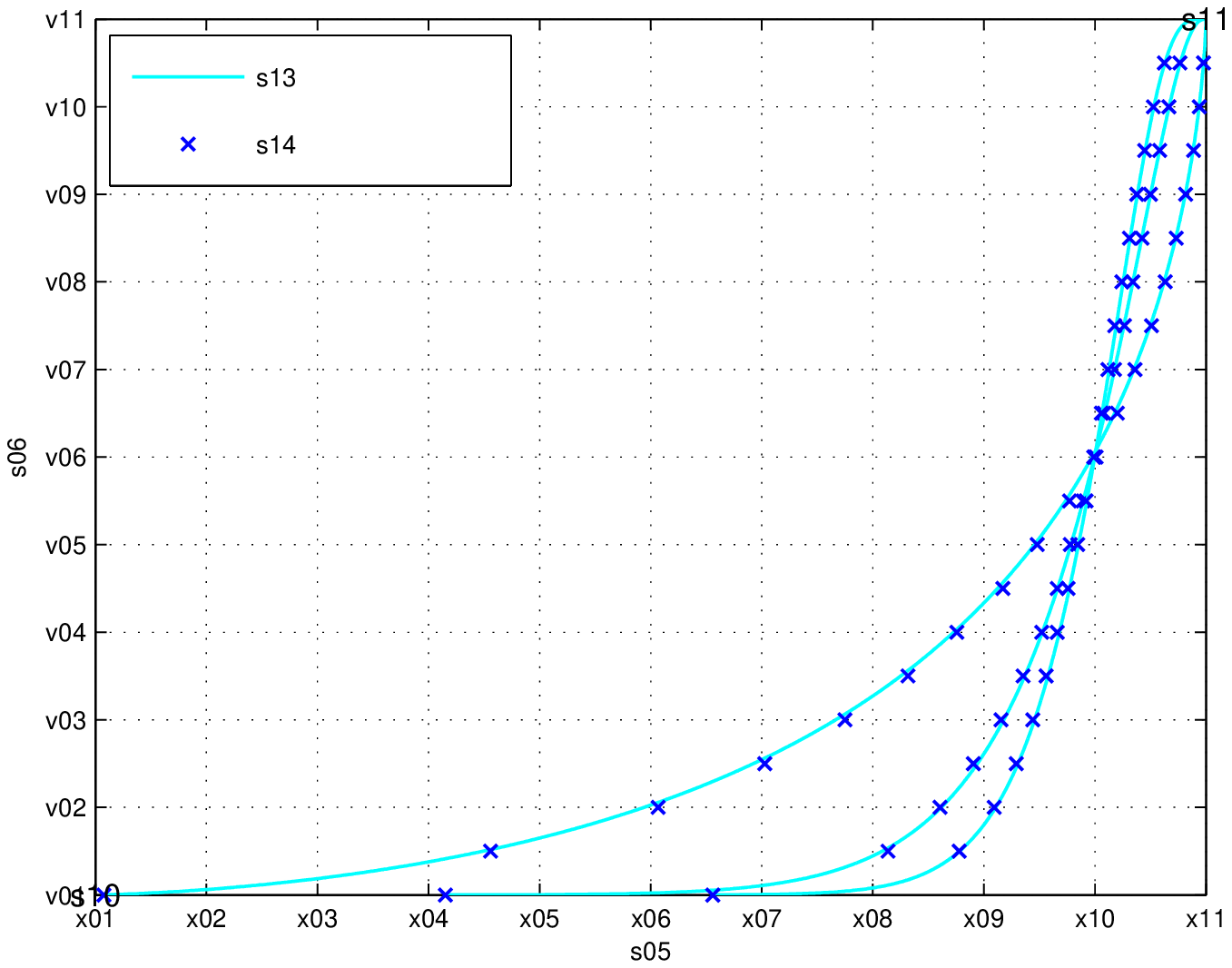}
};
\begin{scope}[x={(image.south east)},y={(image.north west)}]
\draw[black,->] (0.62,0.3) -- (0.82,0.13);
\node[draw=none, font=\scriptsize] at (0.57,0.35) {$\test \in \{1,5,10\} \SI{}{ms}$};
\end{scope}
\end{tikzpicture} 
\quad \label{fig:CDF_pd_test}
}
\hfil
\subfloat[]{
%
%
%
\psfrag{s05}[t][t]{\fontsize{8}{12}\fontseries{m}\mathversion{normal}\fontshape{n}\selectfont \color[rgb]{0,0,0}\setlength{\tabcolsep}{0pt}\begin{tabular}{c}$\pd$\end{tabular}}%
\psfrag{s06}[b][b]{\fontsize{8}{12}\fontseries{m}\mathversion{normal}\fontshape{n}\selectfont \color[rgb]{0,0,0}\setlength{\tabcolsep}{0pt}\begin{tabular}{c}CDF\end{tabular}}%
\psfrag{s10}[][]{\fontsize{10}{15}\fontseries{m}\mathversion{normal}\fontshape{n}\selectfont \color[rgb]{0,0,0}\setlength{\tabcolsep}{0pt}\begin{tabular}{c} \end{tabular}}%
\psfrag{s11}[][]{\fontsize{10}{15}\fontseries{m}\mathversion{normal}\fontshape{n}\selectfont \color[rgb]{0,0,0}\setlength{\tabcolsep}{0pt}\begin{tabular}{c} \end{tabular}}%
\psfrag{s12}[l][l]{\fontsize{8}{12}\fontseries{m}\mathversion{normal}\fontshape{n}\selectfont \color[rgb]{0,0,0}Simulated}%
\psfrag{s13}[l][l]{\fontsize{8}{12}\fontseries{m}\mathversion{normal}\fontshape{n}\selectfont \color[rgb]{0,0,0}Theoretical}%
\psfrag{s14}[l][l]{\fontsize{8}{12}\fontseries{m}\mathversion{normal}\fontshape{n}\selectfont \color[rgb]{0,0,0}Simulated}%
%
\fontsize{8}{12}\fontseries{m}\mathversion{normal}%
\fontshape{n}\selectfont%
%
\psfrag{x01}[t][t]{0}%
\psfrag{x02}[t][t]{0.1}%
\psfrag{x03}[t][t]{0.2}%
\psfrag{x04}[t][t]{0.3}%
\psfrag{x05}[t][t]{0.4}%
\psfrag{x06}[t][t]{0.5}%
\psfrag{x07}[t][t]{0.6}%
\psfrag{x08}[t][t]{0.7}%
\psfrag{x09}[t][t]{0.8}%
\psfrag{x10}[t][t]{0.9}%
\psfrag{x11}[t][t]{1}%
%
\psfrag{v01}[r][r]{0}%
\psfrag{v02}[r][r]{0.1}%
\psfrag{v03}[r][r]{0.2}%
\psfrag{v04}[r][r]{0.3}%
\psfrag{v05}[r][r]{0.4}%
\psfrag{v06}[r][r]{0.5}%
\psfrag{v07}[r][r]{0.6}%
\psfrag{v08}[r][r]{0.7}%
\psfrag{v09}[r][r]{0.8}%
\psfrag{v10}[r][r]{0.9}%
\psfrag{v11}[r][r]{1}%
%
%

\begin{tikzpicture}[scale=1]
\node[anchor=south west,inner sep=0] (image) at (0,0)
{
	\includegraphics[width = \figscale]{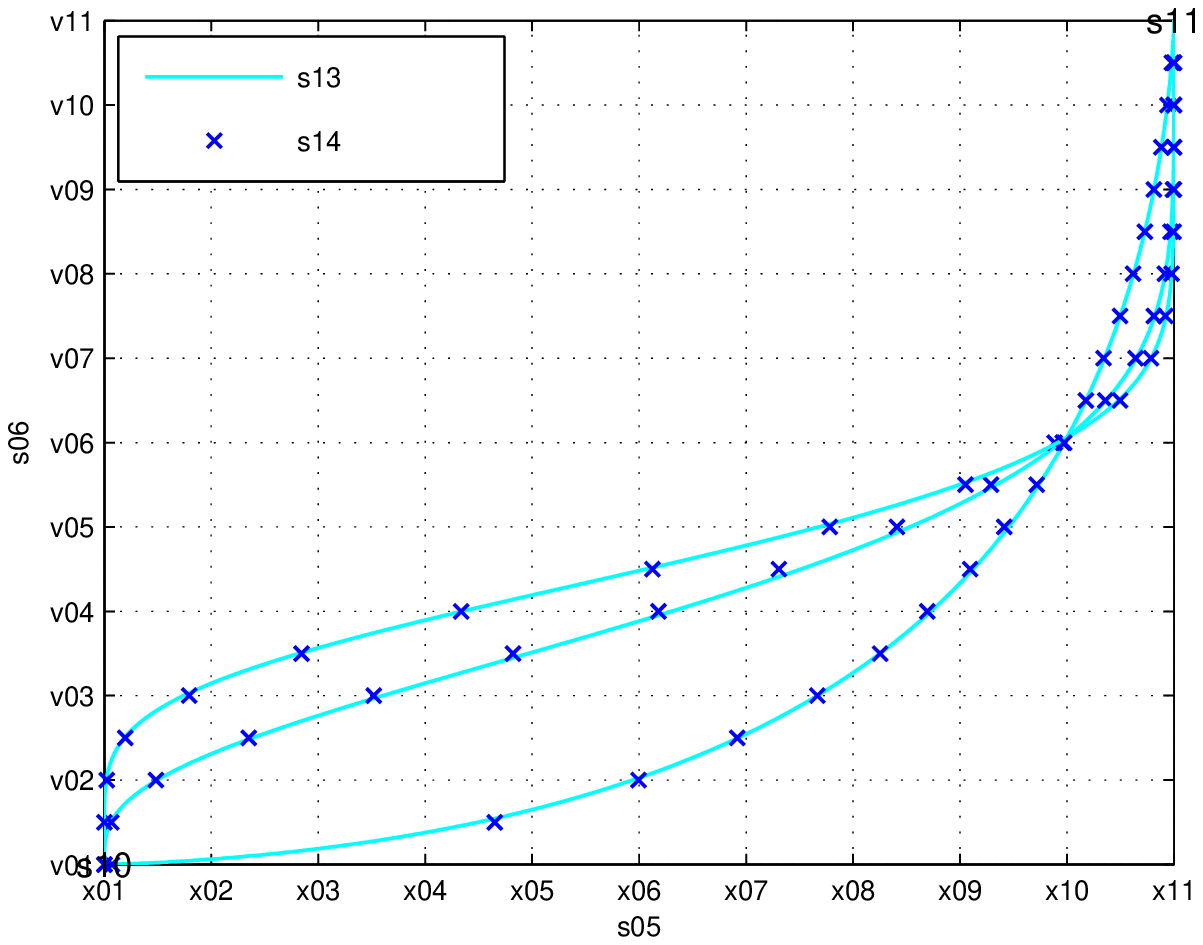}
};
\begin{scope}[x={(image.south east)},y={(image.north west)}]
\draw[black,<-] (0.5,0.45) -- (0.68,0.21); 
\node[draw=none, font=\scriptsize] at (0.73,0.16) {$\tsen \in \{1,5,10\} \SI{}{ms}$};
\end{scope}
\end{tikzpicture}
\quad
\label{fig:CDF_pd_tsen}}
\vspace{0.3cm}
\caption{CDF of $\pd$ for different $\test$ and $\tsen$. (a) $\test \in \{1,5,10\} \SI{}{ms}$ and $\tsen = \SI{1}{ms}$, (b) $\test = \SI{1}{ms}$ and $\tsen \in \{1,5,10\} \SI{}{ms}$.}
\label{fig:CDF_pd}
\vspace{-0.5cm}
\end{figure}

\begin{lemma} \label{lm:lem2}
\normalfont
The cumulative distribution function of $\cz$ is defined as
\begin{align}
\fcz(x) &= \int\limits_{0}^{x} \dcz(t) dt, \label{eq:dis_C0} 
\end{align}
where
\begin{align}
\dcz(x) &= 2^x \ln 2 \frac{(2^x - 1)^{\as - 1}}{\Gamma(\as) \bs^{\as}} \exp\left(-\frac{2^x - 1}{\bs}\right),  \label{eq:den_C0}
\end{align}
and  
\begin{align}
\quad & \as = \frac{\left(\lambdas + |\hs|^2  \right)^2 }{ \lambdas \left(2 \lambdas + 4 |\hs|^2  \right)}   \text{  and  } \nonumber \\ \quad & \bs = \frac{\lambdas \left(2 \lambdas + 4  |\hs|^2 \right)}{\left( \lambdas + |\hs|^2  \right) } \label{eq:para_s}. 
\end{align}
\end{lemma} 
\begin{IEEEproof}
Following the probability density function (pdf) of $|\ehs|^2$ in (\ref{eq:ehs}), the pdf $|\ehs|^2 \frac{\ptranst}{\npo}$ is given by
\begin{align}
\dsnrs(x) &= \lambdasinv \frac{1}{2} \exp\left[-\frac{1}{2} \left( x \lambdas + \ls \right)\right] \times \nonumber \\ \quad  & \left( \frac{x}{\ls} \lambdas \right)^{\frac{\Ks}{4} - \frac{1}{2}} I_{\frac{\Ks}{2}  - 1}  \left( \sqrt{\ls x \lambdas }  \right), \nonumber 
\end{align}
where $I_{(\cdot)}(\cdot)$ represents the modified Bessel function of first kind \cite{grad}. Approximating $\ncchi2(\cdot, \cdot)$ with Gamma distribution $\Gamma(\as, \bs)$ \cite{abramo} gives 
\begin{align}
\dsnrs & \approx \frac{1}{\Gamma(\as)} \frac{x^{\as - 1}}{\bs^{\as}} \exp\left(-\frac{x}{\bs}\right), 
\label{eq:dsnrs}
\end{align} 
where the parameters $\as$ and $\bs$ in (\ref{eq:dsnrs}) are determined by comparing the first two central moments of the two distributions. Finally, by substituting the expression of $\cz$ in (\ref{eq:Cap0}) yields (\ref{eq:den_C0}). 
\end{IEEEproof}

\begin{figure}[!t]

%
%
%
\psfrag{s05}[t][t]{\fontsize{8}{12}\fontseries{m}\mathversion{normal}\fontshape{n}\selectfont \color[rgb]{0,0,0}\setlength{\tabcolsep}{0pt}\begin{tabular}{c}$\text{C}_0$ [bits/sec/Hz]\end{tabular}}%
\psfrag{s06}[b][b]{\fontsize{8}{12}\fontseries{m}\mathversion{normal}\fontshape{n}\selectfont \color[rgb]{0,0,0}\setlength{\tabcolsep}{0pt}\begin{tabular}{c}CDF\end{tabular}}%
\psfrag{s10}[][]{\fontsize{10}{15}\fontseries{m}\mathversion{normal}\fontshape{n}\selectfont \color[rgb]{0,0,0}\setlength{\tabcolsep}{0pt}\begin{tabular}{c} \end{tabular}}%
\psfrag{s11}[][]{\fontsize{10}{15}\fontseries{m}\mathversion{normal}\fontshape{n}\selectfont \color[rgb]{0,0,0}\setlength{\tabcolsep}{0pt}\begin{tabular}{c} \end{tabular}}%
\psfrag{s12}[l][l]{\fontsize{8}{12}\fontseries{m}\mathversion{normal}\fontshape{n}\selectfont \color[rgb]{0,0,0}Simulated}%
\psfrag{s13}[l][l]{\fontsize{8}{12}\fontseries{m}\mathversion{normal}\fontshape{n}\selectfont \color[rgb]{0,0,0}Theoretical}%
\psfrag{s14}[l][l]{\fontsize{8}{12}\fontseries{m}\mathversion{normal}\fontshape{n}\selectfont \color[rgb]{0,0,0}Simulated}%
%
\fontsize{8}{12}\fontseries{m}\mathversion{normal}%
\fontshape{n}\selectfont%
%
\psfrag{x01}[t][t]{0}%
\psfrag{x02}[t][t]{0.5}%
\psfrag{x03}[t][t]{1}%
\psfrag{x04}[t][t]{1.5}%
\psfrag{x05}[t][t]{2}%
\psfrag{x06}[t][t]{2.5}%
\psfrag{x07}[t][t]{3}%
\psfrag{x08}[t][t]{3.5}%
\psfrag{x09}[t][t]{4}%
\psfrag{x10}[t][t]{4.5}%
%
\psfrag{v01}[r][r]{0}%
\psfrag{v02}[r][r]{0.1}%
\psfrag{v03}[r][r]{0.2}%
\psfrag{v04}[r][r]{0.3}%
\psfrag{v05}[r][r]{0.4}%
\psfrag{v06}[r][r]{0.5}%
\psfrag{v07}[r][r]{0.6}%
\psfrag{v08}[r][r]{0.7}%
\psfrag{v09}[r][r]{0.8}%
\psfrag{v10}[r][r]{0.9}%
\psfrag{v11}[r][r]{1}%
%
%

\centering
\begin{tikzpicture}[scale=1]
\node[anchor=south west,inner sep=0] (image) at (0,0)
{
\includegraphics[width= \figscale]{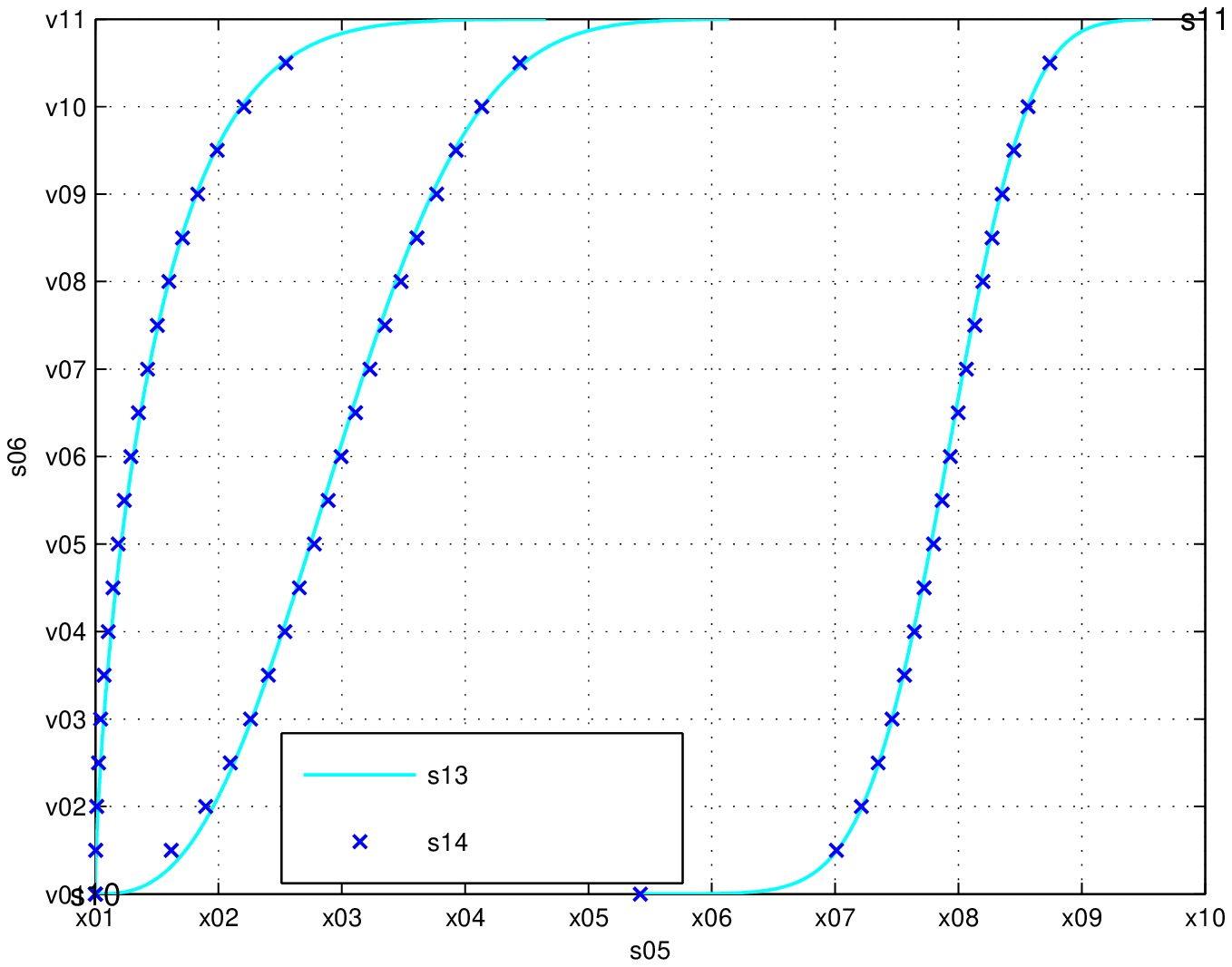}
};
\begin{scope}[x={(image.south east)},y={(image.north west)}]

\draw (0.748,0.52) arc(-250:70:0.04 and 0.02);
\node[draw,fill=gray!10,font=\scriptsize] (text1) at (0.55,0.50) {$\snrso = \SI{10}{dB}$};
\draw[black, <-] (text1.east) -- (0.72,0.50);
\draw (0.31,0.7) arc(-250:70:0.04 and 0.02);
\node[draw,fill=gray!10,font=\scriptsize] (text2) at (0.55,0.68) {$\snrso = \SI{0}{dB}$};
\draw[black, <-] (text2.west) -- (0.365,0.68);
\draw (0.20,0.9) arc(-250:70:0.04 and 0.02);
\node[draw,fill=gray!10,font=\scriptsize] (text3) at (0.55,0.88) {$\snrso = \SI{-10}{dB}$};
\draw[black, <-] (text3.west) -- (0.26,0.88);

\end{scope}
\end{tikzpicture}

\caption{CDF of $\cz$ for different values of $\snrso \in \{-10,0,10 \} \SI{}{dB}$.}
\label{fig:CDF_C0}
\vspace{-0.0cm}
\end{figure}

\begin{lemma} \label{lm:lem3}
\normalfont
The cumulative distribution function of $\co$ is given by  
\begin{align}
\fco(x) &= \int\limits_{0}^{x} \dco(t) dt, \label{eq:dis_C1} 
\end{align}
where
\begin{align}
\dco(x) &= 2^x \ln 2 \frac{(2^x - 1)^{\as - 1} \Gamma(\as + \ap)}{\Gamma(\as) \Gamma(\ap) \bs^{\as} \bp^{\ap}} \left(\frac{1}{\bp} + \frac{2^x - 1}{\bs}\right)^{(\as + \ap)}, \label{eq:den_C1}
\end{align}
and 
\begin{align}
\quad & \ap = \frac{\Kp}{2}  \text{  and  } \bp = \frac{2 \prcvdsr}{\npo \Kp}, \label{eq:para_p} 
\end{align}
where $\as$ and $\bs$ are defined in (\ref{eq:para_s}). 
\end{lemma} 
\begin{IEEEproof}
\tc{See Appendix A.}
\end{IEEEproof}

\begin{figure}[!t]
\centering
\subfloat[]{
%
%
%
\psfrag{s05}[t][t]{\fontsize{8}{12}\fontseries{m}\mathversion{normal}\fontshape{n}\selectfont \color[rgb]{0,0,0}\setlength{\tabcolsep}{0pt}\begin{tabular}{c}$\text{C}_1$ [bits/sec/Hz]\end{tabular}}%
\psfrag{s06}[b][b]{\fontsize{8}{12}\fontseries{m}\mathversion{normal}\fontshape{n}\selectfont \color[rgb]{0,0,0}\setlength{\tabcolsep}{0pt}\begin{tabular}{c}CDF\end{tabular}}%
\psfrag{s10}[][]{\fontsize{10}{15}\fontseries{m}\mathversion{normal}\fontshape{n}\selectfont \color[rgb]{0,0,0}\setlength{\tabcolsep}{0pt}\begin{tabular}{c} \end{tabular}}%
\psfrag{s11}[][]{\fontsize{10}{15}\fontseries{m}\mathversion{normal}\fontshape{n}\selectfont \color[rgb]{0,0,0}\setlength{\tabcolsep}{0pt}\begin{tabular}{c} \end{tabular}}%
\psfrag{s12}[l][l]{\fontsize{8}{12}\fontseries{m}\mathversion{normal}\fontshape{n}\selectfont \color[rgb]{0,0,0}Simulated}%
\psfrag{s13}[l][l]{\fontsize{8}{12}\fontseries{m}\mathversion{normal}\fontshape{n}\selectfont \color[rgb]{0,0,0}Theoretical}%
\psfrag{s14}[l][l]{\fontsize{8}{12}\fontseries{m}\mathversion{normal}\fontshape{n}\selectfont \color[rgb]{0,0,0}Simulated}%
%
\fontsize{8}{12}\fontseries{m}\mathversion{normal}%
\fontshape{n}\selectfont%
%
\psfrag{x01}[t][t]{0}%
\psfrag{x02}[t][t]{0.5}%
\psfrag{x03}[t][t]{1}%
\psfrag{x04}[t][t]{1.5}%
%
\psfrag{v01}[r][r]{0}%
\psfrag{v02}[r][r]{0.1}%
\psfrag{v03}[r][r]{0.2}%
\psfrag{v04}[r][r]{0.3}%
\psfrag{v05}[r][r]{0.4}%
\psfrag{v06}[r][r]{0.5}%
\psfrag{v07}[r][r]{0.6}%
\psfrag{v08}[r][r]{0.7}%
\psfrag{v09}[r][r]{0.8}%
\psfrag{v10}[r][r]{0.9}%
\psfrag{v11}[r][r]{1}%
%
%

\begin{tikzpicture}[scale=1]
\node[anchor=south west,inner sep=0] (image) at (0,0)
{
	\includegraphics[width = \figscale]{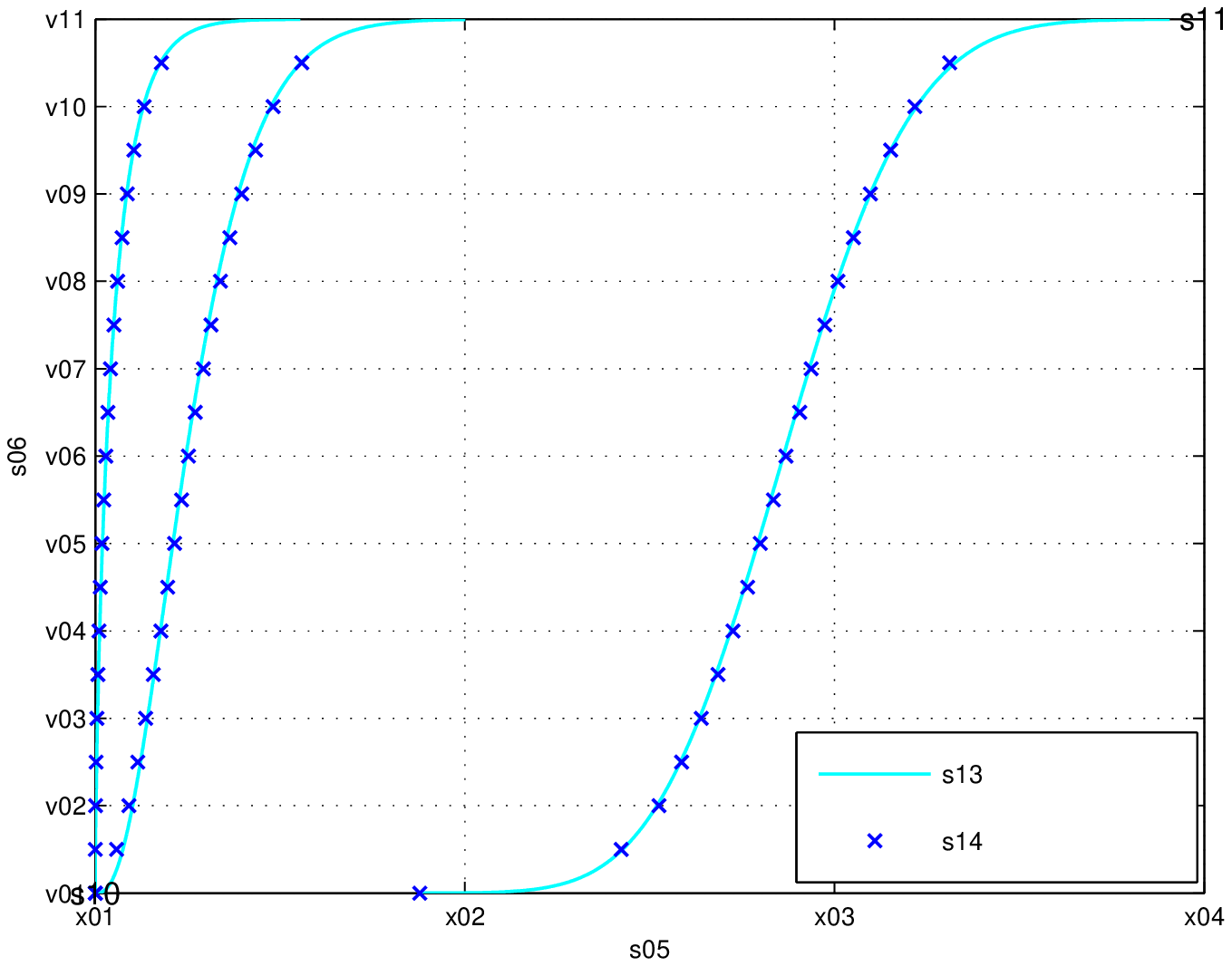} 
};
\begin{scope}[x={(image.south east)},y={(image.north west)}]

\draw (0.618,0.52) arc(-250:70:0.04 and 0.02);
\node[draw,fill=gray!10,font=\scriptsize] (text1) at (0.41,0.50) {$\snrso = \SI{10}{dB}$};
\draw[black, <-] (text1.east) -- (0.592,0.50);
\draw (0.167,0.7) arc(-250:70:0.04 and 0.02);
\node[draw,fill=gray!10,font=\scriptsize] (text2) at (0.41,0.68) {$\snrso = \SI{0}{dB}$};
\draw[black, <-] (text2.west) -- (0.223,0.68);
\draw (0.115,0.9) arc(-250:70:0.04 and 0.02);
\node[draw,fill=gray!10,font=\scriptsize] (text3) at (0.41,0.88) {$\snrso = \SI{-10}{dB}$};
\draw[black, <-] (text3.west) -- (0.175,0.88);

\end{scope}
\end{tikzpicture}

\label{fig:CDF_C1_s}}
\hfil
\subfloat[]{
%
%
%
\psfrag{s05}[t][t]{\fontsize{8}{12}\fontseries{m}\mathversion{normal}\fontshape{n}\selectfont \color[rgb]{0,0,0}\setlength{\tabcolsep}{0pt}\begin{tabular}{c}$\text{C}_1$ [bits/sec/Hz]\end{tabular}}%
\psfrag{s06}[b][b]{\fontsize{8}{12}\fontseries{m}\mathversion{normal}\fontshape{n}\selectfont \color[rgb]{0,0,0}\setlength{\tabcolsep}{0pt}\begin{tabular}{c}CDF\end{tabular}}%
\psfrag{s10}[][]{\fontsize{10}{15}\fontseries{m}\mathversion{normal}\fontshape{n}\selectfont \color[rgb]{0,0,0}\setlength{\tabcolsep}{0pt}\begin{tabular}{c} \end{tabular}}%
\psfrag{s11}[][]{\fontsize{10}{15}\fontseries{m}\mathversion{normal}\fontshape{n}\selectfont \color[rgb]{0,0,0}\setlength{\tabcolsep}{0pt}\begin{tabular}{c} \end{tabular}}%
\psfrag{s12}[l][l]{\fontsize{8}{12}\fontseries{m}\mathversion{normal}\fontshape{n}\selectfont \color[rgb]{0,0,0}Simulated}%
\psfrag{s13}[l][l]{\fontsize{8}{12}\fontseries{m}\mathversion{normal}\fontshape{n}\selectfont \color[rgb]{0,0,0}Theoretical}%
\psfrag{s14}[l][l]{\fontsize{8}{12}\fontseries{m}\mathversion{normal}\fontshape{n}\selectfont \color[rgb]{0,0,0}Simulated}%
%
\fontsize{8}{12}\fontseries{m}\mathversion{normal}%
\fontshape{n}\selectfont%
%
\psfrag{x01}[t][t]{0}%
\psfrag{x02}[t][t]{0.5}%
\psfrag{x03}[t][t]{1}%
\psfrag{x04}[t][t]{1.5}%
\psfrag{x05}[t][t]{2}%
\psfrag{x06}[t][t]{2.5}%
\psfrag{x07}[t][t]{3}%
%
\psfrag{v01}[r][r]{0}%
\psfrag{v02}[r][r]{0.1}%
\psfrag{v03}[r][r]{0.2}%
\psfrag{v04}[r][r]{0.3}%
\psfrag{v05}[r][r]{0.4}%
\psfrag{v06}[r][r]{0.5}%
\psfrag{v07}[r][r]{0.6}%
\psfrag{v08}[r][r]{0.7}%
\psfrag{v09}[r][r]{0.8}%
\psfrag{v10}[r][r]{0.9}%
\psfrag{v11}[r][r]{1}%
%
%

\begin{tikzpicture}[scale=1]
\node[anchor=south west,inner sep=0] (image) at (0,0)
{
	\includegraphics[width = \figscale]{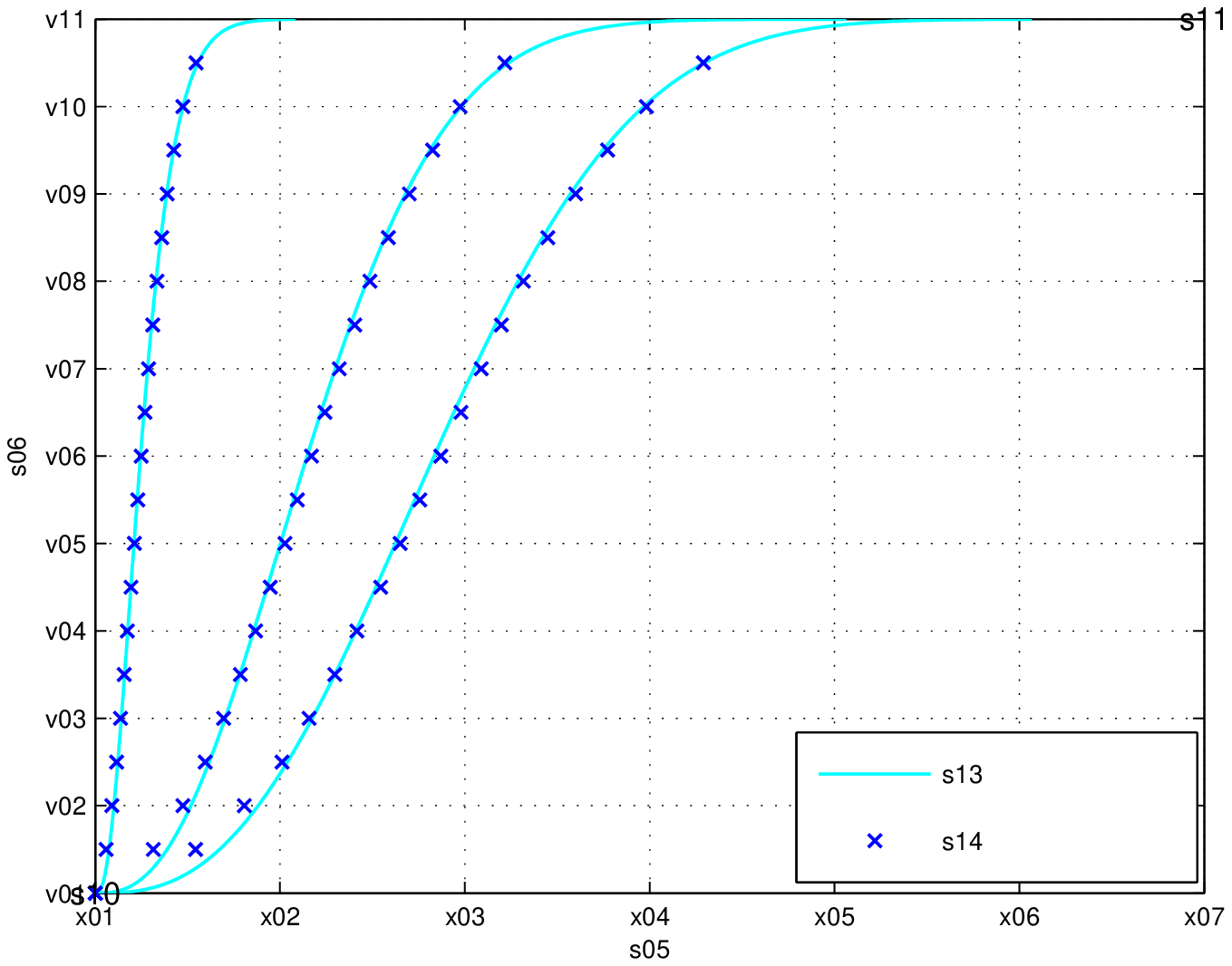} 
};
\begin{scope}[x={(image.south east)},y={(image.north west)}]

\draw (0.34,0.52) arc(-250:70:0.04 and 0.02);
\node[draw,fill=gray!10,font=\scriptsize] (text1) at (0.72,0.5) {$\snrpt = \SI{-10}{dB}$};
\draw[black, <-] (text1.west) -- (0.395,0.5);
\draw (0.287,0.7) arc(-250:70:0.04 and 0.02);
\node[draw,fill=gray!10,font=\scriptsize] (text2) at (0.72,0.68) {$\snrpt = \SI{0}{dB}$};
\draw[black, <-] (text2.west) -- (0.345,0.68);
\draw (0.145,0.9) arc(-250:70:0.04 and 0.02);
\node[draw,fill=gray!10,font=\scriptsize] (text3) at (0.72,0.88) {$\snrpt = \SI{10}{dB}$};
\draw[black, <-] (text3.west) -- (0.205,0.88);

\end{scope}
\end{tikzpicture}
\label{fig:CDF_C1_p2}}
\vspace{0.3cm}
\caption{CDF of $\co$ for different $\snrso$ and $\snrpt$. (a) $\snrso \in \{-10, 0, 10\} \SI{}{dB}$ and $\snrpt = \SI{10}{dB}$, (b) $\snrso = \SI{0}{dB}$ and $\snrpt \in \{-10, 0, 10\} \SI{}{dB}$.}
\label{fig:CDF_C1}
\vspace{-0.5cm}
\end{figure}
The theoretical expressions of the distribution functions depicted in Lemma \ref{lm:lem1}, Lemma \ref{lm:lem2} and Lemma \ref{lm:lem3} are validated by means of simulations in \figurename~\ref{fig:CDF_pd}, \figurename~\ref{fig:CDF_C0} and \figurename~\ref{fig:CDF_C1}, respectively, with different choices of system parameters, these include $\test \in \{1,5,10\} \SI{}{ms}$, $\tsen = \{1,5,10\} \SI{}{ms}$, $\snrso \in \{-10, 0, 10\} \SI{}{dB}$ and  $\snrpt \in \{-10, 0, 10\} \SI{}{dB}$. 

\subsection{Sensing-throughput tradeoff} \label{sec:st_ana}
Here, we establish sensing-throughput tradeoff for the estimation model that includes the estimation time and incorporates variations in the performance parameter. Most importantly, to restrain the harmful interference at the PR due to the variations in the detection probability
, we propose two new PU constraints at the PR, namely, an average constraint and an outage constraint on the detection probability. Based on these constraints, we characterize the sensing-throughput tradeoff for the IS.

\begin{theorem} \label{th:th1}
\normalfont
Subject to an average constraint on $\pd$ at the PR, the sensing-throughput tradeoff is given by  
\begin{align}
\trsac(\ttest, \ttsenac) &= \maxi_{\tc{\test}, \tsen} \e{\pd, \cz, \co}{\rs(\test, \tsen)}, \nonumber \\ 
\quad &= \frac{T- \tsen}{T} \bigg[ \e{\cz}{\cz} (1 - \pfa) \phz + \nonumber \\ \quad & \e{\co}{\co} (1 - \e{\pd}{\pd}) \pho  \bigg], \label{eq:thr_AC} \\
\text{s.t.} & \text{ }  \e{\pd}{\pd} \ge \pdd, \label{eq:AC} \\
\tc{\text{s.t.}} & \text{ }  \tc{0 < \test \le \tsen \le T,} \nonumber
\end{align}
\end{theorem} 
where $\e{\pd}{\cdot}$ represents the expectation with respect to $\pd$, $\e{\pd, \cz, \co}{\cdot}$ denotes the expectation with respect to $\pd$, $\cz$ and $\co$. Unlike (\ref{eq:thr_id_con}), $\pdd$ in (\ref{eq:thr_AC}) represents the constraint on expected detection probability.

\begin{IEEEproof} 
\tc{See Appendix B.
For simplification, the proof of Theorem \ref{th:th1} is included in the proof of Theorem \ref{th:th2}.} 
\end{IEEEproof}

\begin{theorem} \label{th:th2}
\normalfont
Subject to an outage constraint on $\pd$ at the PR, the sensing-throughput tradeoff is given by  
\begin{align}
\trsoc(\ttest, \ttsenoc) &= \maxi_{\tc{\test}, \tsen} \e{\pd, \cz, \co}{\rs(\test, \tsen)}, \nonumber \\ 
\quad &= \frac{T- \tsen}{T} \bigg[ \e{\cz}{\cz} (1 - \pfa) \phz + \nonumber \\ \quad & \e{\co}{\co} (1 - \e{\pd}{\pd}) \pho  \bigg], \label{eq:thr_OC} \\
\text{s.t.} & \text{ }  \p(\pd \le \pdd) \le \mpd, \label{eq:OC} \\
\tc{\text{s.t.}} & \text{ }  \tc{0 < \test \le \tsen \le T,} \nonumber
\end{align}
\end{theorem} 
where $\mpd$ represents the outage constraint. 

\begin{IEEEproof} 
\tc{See Appendix B.}
\end{IEEEproof} 
\tc{In contrast to the ideal model, the sensing-throughput tradeoff investigated by the estimation model (refer to Theorems \ref{th:th1} and \ref{th:th2}) incorporates the imperfect channel knowledge, in this context, the performance characterization considered by the proposed framework are closer to the realistic situations.}

\begin{remark} \label{rem:rem1}
\normalfont
Herein, based on the estimation model, we establish a fundamental relation between estimation time (regulates the variation in the detection probability according to the PU constraint), sensing time (represents the detector performance) and achievable throughput, this relationship is characterized as \textit{estimation-sensing-throughput tradeoff}. Based on this tradeoff, we determine the suitable estimation $\test = \ttest$ and sensing time $\tsen = \ttsen$ that attains a maximum achievable throughput $\trs(\ttest,\ttsen)$ for the IS.  
\end{remark}

\begin{coro} \label{cor:cor1}
\normalfont
\tc{Theorems \ref{th:th1} and \ref{th:th2} consider the optimization of the average throughput to incorporate the effect of variations due to channels estimation, and subsequently determine the suitable sensing and the suitable estimation time. Here, we investigate an alternative approach to the optimization problem described in (\ref{eq:thr_id}) to capture these variations, whereby for a certain estimation time $\test$, the suitable sensing time subject to the average constraint is determined as} 
\tc{
\begin{align}
\ttsen &= \argmaxi_{\tsen} \rs(\test, \tsen), \label{eq:C_sen_AC} \\ 
\quad &= \frac{T- \tsen}{T} \bigg[ \cz (1 - \pfa) \phz + \co (1 - \pd) \pho  \bigg], \nonumber \\
\text{s.t.} & \text{ }  \e{\pd}{\pd} \ge \pdd, \nonumber \\
\tc{\text{s.t.}} & \text{ }  \tc{0 < \test \le \tsen \le T.} \nonumber
\end{align}
}
\tc{
Similarly, the suitable sensing time subject to the outage constraint is determined as}
\tc{
\begin{align}
\ttsen &= \argmaxi_{\tsen} \rs(\test, \tsen), \label{eq:C_sen_OC} \\ 
\quad &= \frac{T- \tsen}{T} \bigg[ \cz (1 - \pfa) \phz + \co (1 - \pd) \pho  \bigg], \nonumber \\
\text{s.t.} & \text{ }   \p(\pd \le \pdd) \le \mpd, \nonumber \\
\tc{\text{s.t.}} & \text{ }  \tc{0 < \test \le \tsen \le T.} \nonumber
\end{align}
}
\tc{
In contrast to (\ref{eq:thr_AC}) and (\ref{eq:thr_OC}), the suitable sensing time evaluated in (\ref{eq:C_sen_AC}) and (\ref{eq:C_sen_OC}) entails the variations due to channel estimation. Hence, the secondary throughput subject to the average and the outage constraints captures the variations in the suitable sensing time and the performance parameters is determined as} 
\tc{
\begin{align}
\e{\pd, \cz, \co, \ttsen}{\rs(\test, \ttsen)} \label{eq:C_thr},
\end{align}
} 
\tc{
where $\e{\pd, \cz, \co, \ttsen}{\cdot}$ corresponds to an expection over $\pd, \cz, \co, \ttsen$.
Following Remark \ref{rem:rem1}, we further optimize the average throughput, defined in (\ref{eq:C_thr}), over the estimation time} 
\tc{
\begin{align}
\trs(\ttest, \ttsenac) &= \maxi_{\test} \e{\pd, \cz, \co, \ttsen}{\rs(\test, \ttsen)} \label{eq:C_thr_AC}. 
\end{align}
}
\tc{
In this way, we establish an estimation-sensing-throughput tradeoff for the alternative approach to determine the suitable estimation time.    
}
\end{coro}
\begin{remark} \label{rm:rem2} 
\normalfont
\tc{Complementing the analysis in \cite{Liang08}, it is complicated to obtain a closed-form expression of $\ttsen$, thereby rendering the analytical tractability of its distribution function difficult. In view of this, we capture the performance of the alternative approach by means of simulations.}
\end{remark} 
\section{Numerical Results} \label{sec:num_ana}
Here, we investigate the performance of the IS based on the proposed approach. To accomplish this: (i) we perform simulations to validate the expressions obtained, (ii) we analyze the performance loss incurred due to the estimation. In this regard, we consider the ideal model to benchmark and evaluate the performance loss, (iii) we establish mathematical justification to the considered approximations. Although the expressions derived in this paper depicting the sensing-throughput analysis are general and applicable to all CR systems, the parameters are selected in such a way that they closely relate to the deployment scenario described in \figurename~{\ref{fig:scenario}. Unless stated explicitly, the choice of the parameters given in Table \ref{tb:tb2} is considered for the analysis. 
\begin{table}
\renewcommand{\arraystretch}{1.4}
\caption{Parameters for Numerical Analysis}
\label{tb:tb2}
\centering
\scriptsize{
\begin{tabular}{c||c}
\hline
\bfseries Parameter & \bfseries Value \\
\hline\hline
$\fsam$  & $\SI{1}{MHz}$ \\ \hline
$\phpo, \phpt$ & $\SI{-100}{dB}$ \\ \hline
$\phs$ & $\SI{-80}{dB}$ \\ \hline 
$T$ & $\SI{100}{ms}$ \\ \hline 
$\pdd$ & 0.9 \\ \hline 
$\mpd$ & $0.05$ \\ \hline 
$\npo$ & $\SI{-100}{dBm}$ \\ \hline
$\snrrcvd$ & $\SI{-10}{dB}$ \\ \hline
$\snrpt$ & $\SI{-10}{dB}$ \\ \hline
$\snrso$ & $\SI{10}{dB}$ \\ \hline
$\spo = \ptranpt$ & $-\SI{10}{dBm}$ \\ \hline
$\ptranst$ & $-\SI{10}{dBm}$ \\ \hline
$\pho = 1 - \phz$ & 0.2 \\ \hline
$\test$ & $\SI{5}{ms}$ \\ \hline
$\Ks$ & 10 \\ \hline 
$\Kp$ & $1000$ \\ \hline
\end{tabular}}
\end{table}


\begin{figure}[!t]
\centering
%
%
%
\psfrag{s08}[b][b]{\fontsize{8}{12}\fontseries{m}\mathversion{normal}\fontshape{n}\selectfont \color[rgb]{0,0,0}\setlength{\tabcolsep}{0pt}\begin{tabular}{c}$\rs(\test = \SI{5}{ms}, \tsen)$ [bits/sec/Hz]\end{tabular}}%
\psfrag{s09}[t][t]{\fontsize{8}{12}\fontseries{m}\mathversion{normal}\fontshape{n}\selectfont \color[rgb]{0,0,0}\setlength{\tabcolsep}{0pt}\begin{tabular}{c}$\tsen$ [ms]\end{tabular}}%
\psfrag{s13}[][]{\fontsize{10}{15}\fontseries{m}\mathversion{normal}\fontshape{n}\selectfont \color[rgb]{0,0,0}\setlength{\tabcolsep}{0pt}\begin{tabular}{c} \end{tabular}}%
\psfrag{s14}[][]{\fontsize{10}{15}\fontseries{m}\mathversion{normal}\fontshape{n}\selectfont \color[rgb]{0,0,0}\setlength{\tabcolsep}{0pt}\begin{tabular}{c} \end{tabular}}%
\psfrag{s15}[l][l]{\fontsize{8}{12}\fontseries{m}\mathversion{normal}\fontshape{n}\selectfont \color[rgb]{0,0,0}Simulated}%
\psfrag{s16}[l][l]{\fontsize{8}{12}\fontseries{m}\mathversion{normal}\fontshape{n}\selectfont \color[rgb]{0,0,0}IM}%
\psfrag{s17}[l][l]{\fontsize{8}{12}\fontseries{m}\mathversion{normal}\fontshape{n}\selectfont \color[rgb]{0,0,0}EM-AC, Thm. 1}%
\psfrag{s18}[l][l]{\fontsize{8}{12}\fontseries{m}\mathversion{normal}\fontshape{n}\selectfont \color[rgb]{0,0,0}EM-OC, Thm. 2}%
\psfrag{s19}[l][l]{\fontsize{8}{12}\fontseries{m}\mathversion{normal}\fontshape{n}\selectfont \color[rgb]{0,0,0}$\trs(\test, \ttsen)$}%
\psfrag{s20}[l][l]{\fontsize{8}{12}\fontseries{m}\mathversion{normal}\fontshape{n}\selectfont \color[rgb]{0,0,0}Simulated}%
\psfrag{s21}[b][b]{\fontsize{8}{12}\fontseries{m}\mathversion{normal}\fontshape{n}\selectfont \color[rgb]{0,0,0}\setlength{\tabcolsep}{0pt}\begin{tabular}{c}Zoom\end{tabular}}%
%
\fontsize{8}{12}\fontseries{m}\mathversion{normal}%
\fontshape{n}\selectfont%
%
\psfrag{x01}[t][t]{5}%
\psfrag{x02}[t][t]{5.2}%
\psfrag{x03}[t][t]{5.4}%
\psfrag{x04}[t][t]{5.6}%
\psfrag{x05}[t][t]{1}%
\psfrag{x06}[t][t]{2}%
\psfrag{x07}[t][t]{3}%
\psfrag{x08}[t][t]{4}%
\psfrag{x09}[t][t]{5}%
\psfrag{x10}[t][t]{6}%
\psfrag{x11}[t][t]{7}%
\psfrag{x12}[t][t]{8}%
\psfrag{x13}[t][t]{9}%
\psfrag{x14}[t][t]{10}%
%
\psfrag{v01}[r][r]{2.55}%
\psfrag{v02}[r][r]{2.6}%
\psfrag{v03}[r][r]{2.65}%
\psfrag{v04}[r][r]{0}%
\psfrag{v05}[r][r]{0.5}%
\psfrag{v06}[r][r]{1}%
\psfrag{v07}[r][r]{1.5}%
\psfrag{v08}[r][r]{2}%
\psfrag{v09}[r][r]{2.5}%
%
%

\begin{tikzpicture}[scale=1]
\node[anchor=south west,inner sep=0] (image) at (0,0)
{
        \includegraphics[width= \figscale]{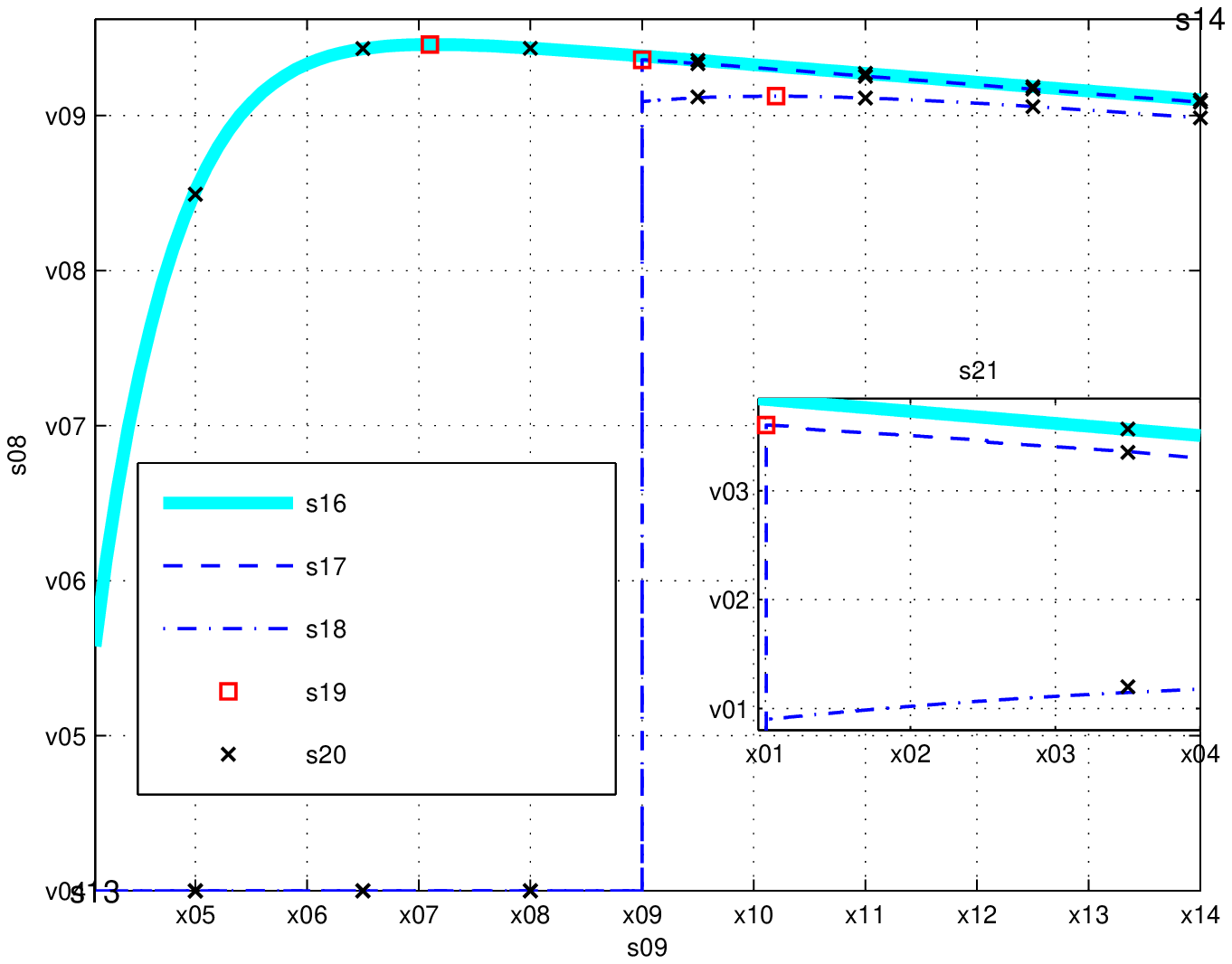}
};
\begin{scope}[x={(image.south east)},y={(image.north west)}]


\draw[black,thick,<->] (0.088,0.1) --  node[above, font=\small] {$\test$} (0.512,0.1);

\end{scope}
\end{tikzpicture}
\caption{\tc{Sensing-throughput tradeoff for the ideal model (IM) and estimation model (EM), $\snrrcvd = \SI{-10}{dB}$, $\test = \SI{5}{ms}$ and $\mpd = 0.05$.}}
\label{fig:ST_gen}
\vspace{-0.0cm}
\end{figure}

At first, we analyze the performance of the IS in terms of sensing-throughput tradeoff corresponding to the ideal model (IM) and estimation model (EM) by fixing $\test = \SI{5}{ms}$, \tc{refer to} \figurename~\ref{fig:ST_gen}. In contrast to constraint on $\pd$ for the ideal model, we employ average constraint (EM-AC) and outage constraint (EM-OC) for the proposed estimation model. With the inclusion of received power based estimation in the frame structure, the ST achieves no throughput at the SR for the interval $\test$. For the given cases, namely, IM, EM-AC and EM-OC, a suitable sensing time that results in a maximum throughput $\trs(\test = \SI{5}{ms},\ttsen)$ is determined. Apart form that, a performance degradation is depicted in terms of the achievable throughput, refer to \figurename~\ref{fig:ST_gen}. For $\mpd = 0.05$, it is observed that the outage constraint is more sensitive to the performance loss in comparison to average constraint. It is clear that the analysis illustrated in \figurename~\ref{fig:ST_gen} is obtained for a certain choice of system parameters, particularly $\snrrcvd = -\SI{10}{dB}$, $\test = \SI{5}{ms}$ and $\mpd = 0.05$. To acquire more insights, we consider the effect of variation of these parameters on the performance of IS, subsequently. 

\begin{figure}[!t]

%
%
%
\psfrag{s05}[b][b]{\fontsize{8}{12}\fontseries{m}\mathversion{normal}\fontshape{n}\selectfont \color[rgb]{0,0,0}\setlength{\tabcolsep}{0pt}\begin{tabular}{c}$\trs(\test = \SI{5}{ms},\ttsen)$ [bits/sec/Hz]\end{tabular}}%
\psfrag{s06}[t][t]{\fontsize{8}{12}\fontseries{m}\mathversion{normal}\fontshape{n}\selectfont \color[rgb]{0,0,0}\setlength{\tabcolsep}{0pt}\begin{tabular}{c}$\snrrcvd$ [dB]\end{tabular}}%
\psfrag{s10}[][]{\fontsize{10}{15}\fontseries{m}\mathversion{normal}\fontshape{n}\selectfont \color[rgb]{0,0,0}\setlength{\tabcolsep}{0pt}\begin{tabular}{c} \end{tabular}}%
\psfrag{s11}[][]{\fontsize{10}{15}\fontseries{m}\mathversion{normal}\fontshape{n}\selectfont \color[rgb]{0,0,0}\setlength{\tabcolsep}{0pt}\begin{tabular}{c} \end{tabular}}%
\psfrag{s12}[l][l]{\fontsize{8}{12}\fontseries{m}\mathversion{normal}\fontshape{n}\selectfont \color[rgb]{0,0,0}EM-OC, Thm. 2}%
\psfrag{s13}[l][l]{\fontsize{8}{12}\fontseries{m}\mathversion{normal}\fontshape{n}\selectfont \color[rgb]{0,0,0}IM}%
\psfrag{s14}[l][l]{\fontsize{8}{12}\fontseries{m}\mathversion{normal}\fontshape{n}\selectfont \color[rgb]{0,0,0}EM-AC, Thm. 1}%
\psfrag{s15}[l][l]{\fontsize{8}{12}\fontseries{m}\mathversion{normal}\fontshape{n}\selectfont \color[rgb]{0,0,0}EM-OC, Thm. 2}%
%
\fontsize{8}{12}\fontseries{m}\mathversion{normal}%
\fontshape{n}\selectfont%
%
\psfrag{x01}[t][t]{-15}%
\psfrag{x02}[t][t]{-10}%
\psfrag{x03}[t][t]{-5}%
\psfrag{x04}[t][t]{0}%
\psfrag{x05}[t][t]{5}%
%
\psfrag{v01}[r][r]{0.5}%
\psfrag{v02}[r][r]{1}%
\psfrag{v03}[r][r]{1.5}%
\psfrag{v04}[r][r]{2}%
\psfrag{v05}[r][r]{2.5}%
%
%

\centering
\includegraphics[width= \figscale]{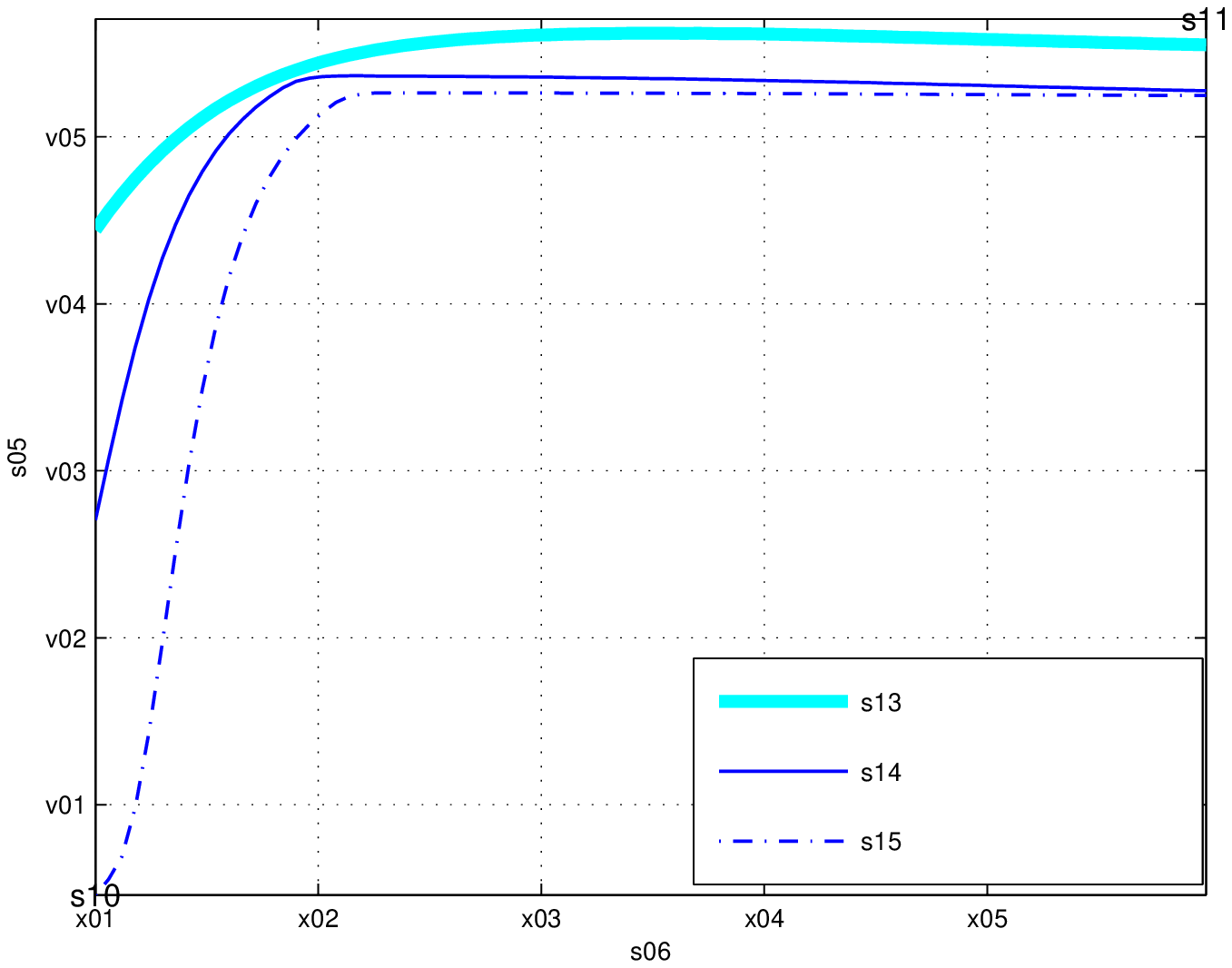}
\caption{\tc{Achievable throughput versus the $\snrrcvd$ with $\tau\sub{est} = \SI{5}{ms}$.}}
\label{fig:optT_snr}
\vspace{-0.7cm}
\end{figure}

Hereafter, for the analysis, we consider the theoretical expressions and choose to operate at a suitable sensing time. Next, we capture the variation in the achievable throughput against the received signal to noise ratio $\snrrcvd$ at the ST with $\test = \SI{5}{ms}$, \tc{refer to} \figurename~\ref{fig:optT_snr}. For $\snrrcvd < -\SI{10}{dB}$, the estimation model incurs a significant performance loss. This clearly reveals that the ideal model overestimates the performance of IS. \tc{From the previous discussion, it is concluded that the inclusion of average and outage constraints (depicted by the proposed framework) preclude the excessive interference at the PR arising due to channel estimation without considerably degrading the performance of the IS.}

\begin{figure}[!t]
\centering
\subfloat[]{
%
%
%
\psfrag{s02}[b][b]{\fontsize{8}{12}\fontseries{m}\mathversion{normal}\fontshape{n}\selectfont \color[rgb]{0,0,0}\setlength{\tabcolsep}{0pt}\begin{tabular}{c}$\trs(\test,\tsen)$ [bits/sec/Hz]\end{tabular}}%
\psfrag{s03}[lt][lt]{\fontsize{8}{12}\fontseries{m}\mathversion{normal}\fontshape{n}\selectfont \color[rgb]{0,0,0}\setlength{\tabcolsep}{0pt}\begin{tabular}{l}$\tsen$ [ms]\end{tabular}}%
\psfrag{s04}[rt][rt]{\fontsize{8}{12}\fontseries{m}\mathversion{normal}\fontshape{n}\selectfont \color[rgb]{0,0,0}\setlength{\tabcolsep}{0pt}\begin{tabular}{r}$\test$ [ms]\end{tabular}}%
%
\fontsize{8}{12}\fontseries{m}\mathversion{normal}%
\fontshape{n}\selectfont%
%
\psfrag{x01}[t][t]{0}%
\psfrag{x02}[t][t]{5}%
\psfrag{x03}[t][t]{10}%
\psfrag{x04}[t][t]{15}%
\psfrag{x05}[t][t]{20}%
\psfrag{x06}[t][t]{25}%
%
\psfrag{v01}[r][r]{0}%
\psfrag{v02}[r][r]{5}%
\psfrag{v03}[r][r]{10}%
%
\psfrag{z01}[r][r]{0}%
\psfrag{z02}[r][r]{0.5}%
\psfrag{z03}[r][r]{1}%
\psfrag{z04}[r][r]{1.5}%
\psfrag{z05}[r][r]{2}%
\psfrag{z06}[r][r]{2.5}%
\psfrag{z07}[r][r]{3}%
%
%

\centering
\begin{tikzpicture}[scale=1]
\node[anchor=south west,inner sep=0] (image) at (0,0)
{
\includegraphics[width= \figscale]{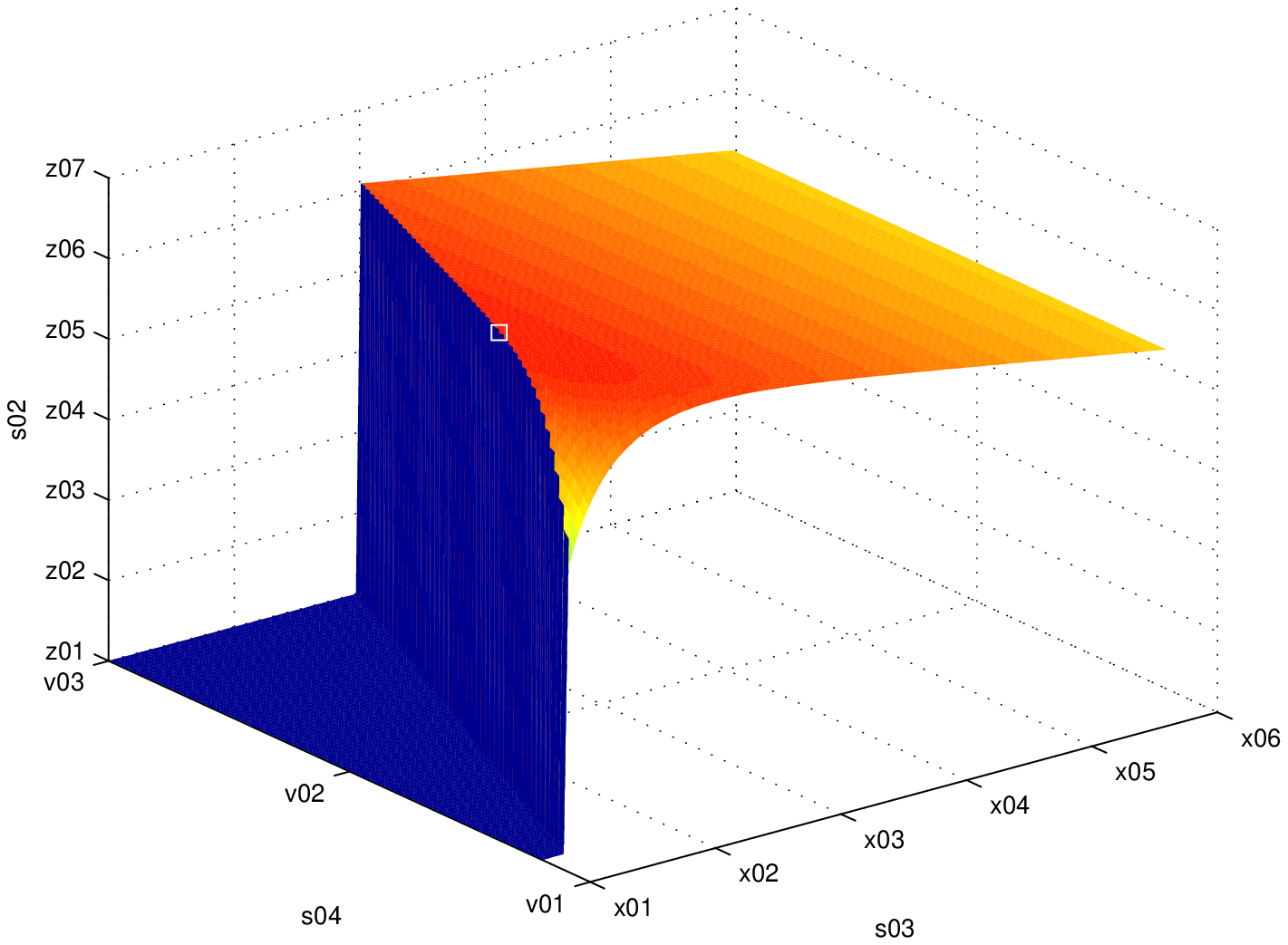}
};
\begin{scope}[x={(image.south east)},y={(image.north west)}]
\draw[black,->] (0.389,0.83) -- (0.389,0.656);
\node[draw=none, font=\scriptsize] at (0.389, 0.86) {$\rs(\ttest, \ttsen)$};
 
\end{scope}
\end{tikzpicture}

\label{fig:EST_ac}}
\hfil
\subfloat[]{
%
%
%
\psfrag{s02}[b][b]{\fontsize{8}{12}\fontseries{m}\mathversion{normal}\fontshape{n}\selectfont \color[rgb]{0,0,0}\setlength{\tabcolsep}{0pt}\begin{tabular}{c}$\trs(\test,\tsen)$ [bits/sec/Hz]\end{tabular}}%
\psfrag{s03}[lt][lt]{\fontsize{8}{12}\fontseries{m}\mathversion{normal}\fontshape{n}\selectfont \color[rgb]{0,0,0}\setlength{\tabcolsep}{0pt}\begin{tabular}{l}$\tsen$ [ms]\end{tabular}}%
\psfrag{s04}[rt][rt]{\fontsize{8}{12}\fontseries{m}\mathversion{normal}\fontshape{n}\selectfont \color[rgb]{0,0,0}\setlength{\tabcolsep}{0pt}\begin{tabular}{r}$\test$ [ms]\end{tabular}}%
%
\fontsize{8}{12}\fontseries{m}\mathversion{normal}%
\fontshape{n}\selectfont%
%
\psfrag{x01}[t][t]{0}%
\psfrag{x02}[t][t]{5}%
\psfrag{x03}[t][t]{10}%
\psfrag{x04}[t][t]{15}%
\psfrag{x05}[t][t]{20}%
\psfrag{x06}[t][t]{25}%
%
\psfrag{v01}[r][r]{0}%
\psfrag{v02}[r][r]{5}%
\psfrag{v03}[r][r]{10}%
%
\psfrag{z01}[r][r]{0}%
\psfrag{z02}[r][r]{0.5}%
\psfrag{z03}[r][r]{1}%
\psfrag{z04}[r][r]{1.5}%
\psfrag{z05}[r][r]{2}%
\psfrag{z06}[r][r]{2.5}%
\psfrag{z07}[r][r]{3}%
%
%

\centering
\begin{tikzpicture}[scale=1]
\node[anchor=south west,inner sep=0] (image) at (0,0)
{
\includegraphics[width= \figscale]{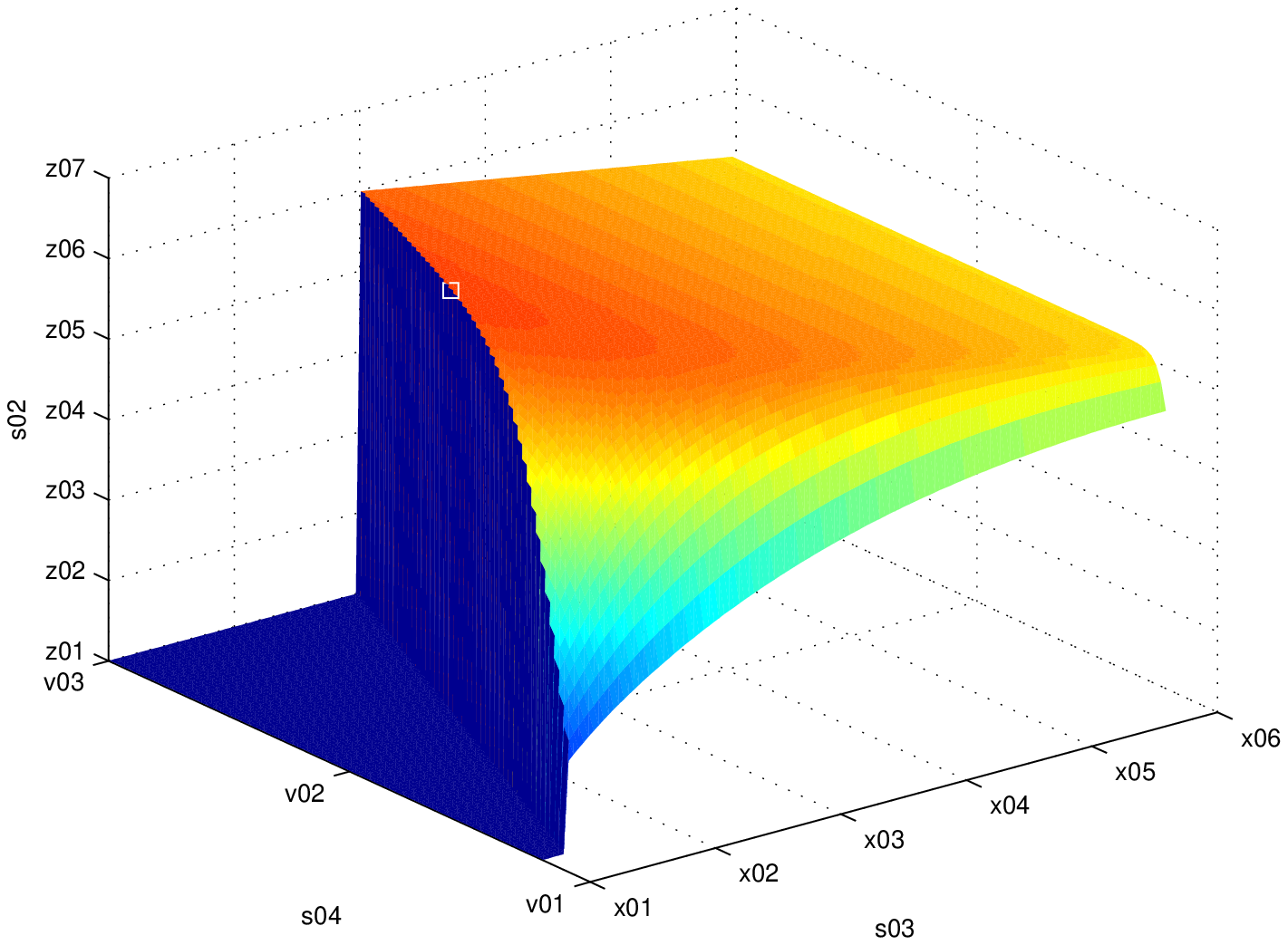}
};
\begin{scope}[x={(image.south east)},y={(image.north west)}]
\draw[black,->] (0.353,0.83) -- (0.353,0.699);
\node[draw=none, font=\scriptsize] at (0.353, 0.86) {$\rs(\ttest, \ttsen)$};
 
\end{scope}
\end{tikzpicture}
\label{fig:EST_oc}}
\vspace{4mm}
\caption{\tc{Estimation-sensing-throughput tradeoff for the estimation model for (a) average constraint and (b) outage constraint with $\mpd = 0.05$.}}
\label{fig:EST}
\end{figure}

Upon maximizing the secondary throughput, it is interesting to analyze the variation of the achievable throughput with the estimation time. Corresponding to the estimation model, \figurename~\ref{fig:EST} illustrates a tradeoff among the estimation time, the sensing time and the throughput, \tc{refer to} Remark \ref{rem:rem1}. \tc{From \figurename~\ref{fig:EST}, it can be noticed that the function $\rs(\test, \tsen)$ is well-behaved in the region $0 < \test \le \tsen \le T$ and consists of a global maximum. This tradeoff depicted by the proposed framework, presented in \figurename~\ref{fig:optT_test}, can be explained from the fact that low values of estimation time result in large variations in $\pd$.} To counteract and satisfy the average and the outage constraints, the corresponding thresholds shift to a lower value. This causes an increase in $\pfa$, thereby increasing the sensing-throughput curvature. As a result, the suitable sensing time is obtained at a higher value. However, beyond a certain value ($\ttest$), a further increase in estimation time slightly contributes to performance improvement and largely consumes the time resources. As a consequence to the estimation-sensing-throughput tradeoff, we determine the suitable estimation time that yields an achievable throughput $\trs(\ttest,\ttsen)$. 

\begin{figure}[!t]

%
%
%
\psfrag{s05}[b][b]{\fontsize{8}{12}\fontseries{m}\mathversion{normal}\fontshape{n}\selectfont \color[rgb]{0,0,0}\setlength{\tabcolsep}{0pt}\begin{tabular}{c}$\trs(\test,\ttsen)$ [bits/sec/Hz]\end{tabular}}%
\psfrag{s06}[t][t]{\fontsize{8}{12}\fontseries{m}\mathversion{normal}\fontshape{n}\selectfont \color[rgb]{0,0,0}\setlength{\tabcolsep}{0pt}\begin{tabular}{c}$\test$ [ms]\end{tabular}}%
\psfrag{s10}[][]{\fontsize{10}{15}\fontseries{m}\mathversion{normal}\fontshape{n}\selectfont \color[rgb]{0,0,0}\setlength{\tabcolsep}{0pt}\begin{tabular}{c} \end{tabular}}%
\psfrag{s11}[][]{\fontsize{10}{15}\fontseries{m}\mathversion{normal}\fontshape{n}\selectfont \color[rgb]{0,0,0}\setlength{\tabcolsep}{0pt}\begin{tabular}{c} \end{tabular}}%
\psfrag{s12}[l][l]{\fontsize{8}{12}\fontseries{m}\mathversion{normal}\fontshape{n}\selectfont \color[rgb]{0,0,0}$\trs(\ttest,\ttsen)$}%
\psfrag{s13}[l][l]{\fontsize{8}{12}\fontseries{m}\mathversion{normal}\fontshape{n}\selectfont \color[rgb]{0,0,0}IM}%
\psfrag{s14}[l][l]{\fontsize{8}{12}\fontseries{m}\mathversion{normal}\fontshape{n}\selectfont \color[rgb]{0,0,0}EM-AC, Thm. 1}%
\psfrag{s15}[l][l]{\fontsize{8}{12}\fontseries{m}\mathversion{normal}\fontshape{n}\selectfont \color[rgb]{0,0,0}EM-OC, Thm. 2}%
\psfrag{s16}[l][l]{\fontsize{8}{12}\fontseries{m}\mathversion{normal}\fontshape{n}\selectfont \color[rgb]{0,0,0}Corollary 1}%
\psfrag{s17}[l][l]{\fontsize{8}{12}\fontseries{m}\mathversion{normal}\fontshape{n}\selectfont \color[rgb]{0,0,0}$\trs(\ttest,\ttsen)$}%
%
\fontsize{8}{12}\fontseries{m}\mathversion{normal}%
\fontshape{n}\selectfont%
%
\psfrag{x01}[t][t]{1}%
\psfrag{x02}[t][t]{2}%
\psfrag{x03}[t][t]{3}%
\psfrag{x04}[t][t]{4}%
\psfrag{x05}[t][t]{5}%
\psfrag{x06}[t][t]{6}%
\psfrag{x07}[t][t]{7}%
\psfrag{x08}[t][t]{8}%
\psfrag{x09}[t][t]{9}%
\psfrag{x10}[t][t]{10}%
%
\psfrag{v01}[r][r]{1.8}%
\psfrag{v02}[r][r]{1.9}%
\psfrag{v03}[r][r]{2}%
\psfrag{v04}[r][r]{2.1}%
\psfrag{v05}[r][r]{2.2}%
\psfrag{v06}[r][r]{2.3}%
\psfrag{v07}[r][r]{2.4}%
\psfrag{v08}[r][r]{2.5}%
\psfrag{v09}[r][r]{2.6}%
\psfrag{v10}[r][r]{2.7}%
%
%

\centering
\begin{tikzpicture}[scale=1]
\node[anchor=south west,inner sep=0] (image) at (0,0)
{
\includegraphics[width= \figscale]{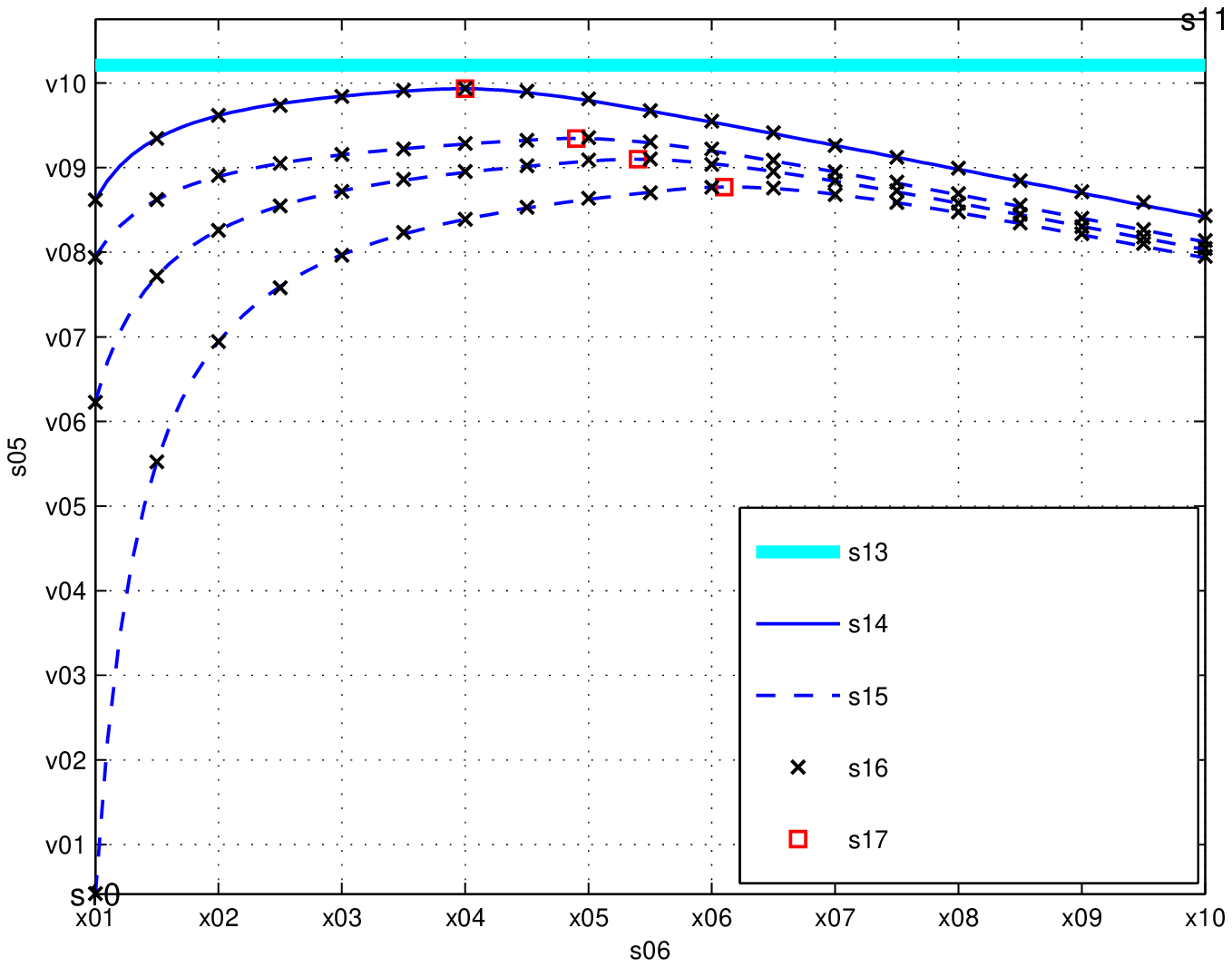}
};
\begin{scope}[x={(image.south east)},y={(image.north west)}]
\draw[black,->] (0.25,0.64) -- (0.18,0.84);
\node[draw=none, font=\scriptsize] at (0.35, 0.58) {$\mpd \in \{0.05,0.10,0.15\}$};
 

\end{scope}
\end{tikzpicture}

\caption{\tc{Estimation-sensing-throughput tradeoff for the average and the outage constraints with $\snrrcvd = \SI{-10}{dB}$, where the throughput is maximized over the sensing time, $\trs(\test,\ttsen)$. The estimation-sensing-throughput tradeoff is utilized to determine a suitable estimation time $\ttest$ that maximizes the throughput, $\trs(\ttest,\ttsen)$.}}
\label{fig:optT_test}
\end{figure}

Besides that, we consider the variation in the achievable throughput for different values of the outage constraint, refer to \figurename~\ref{fig:optT_test}. It is observed that for the selected choice of $\mpd$, the outage constraint is severe as compared to the average constraint, hence, results in a lower throughput. Thus, depending on the nature of policy (aggressive or conservative) followed by the regulatory bodies towards the interference at the primary system, it is possible to define $\mpd$ accordingly during the system design. \tc{Moreover, it is observed that the alternative approach proposed in Corollary \ref{cor:cor1} does not present any noticeable performance difference depicted in terms of the achievable throughput corresponding to the one characterized in the Theorems \ref{th:th1} and \ref{th:th2}. 
}
\begin{figure}[!t]
\centering
\subfloat[]{
%
%
%
\psfrag{s05}[b][b]{\fontsize{8}{12}\fontseries{m}\mathversion{normal}\fontshape{n}\selectfont \color[rgb]{0,0,0}\setlength{\tabcolsep}{0pt}\begin{tabular}{c}$\e{\pd}{\pd}$\end{tabular}}%
\psfrag{s06}[t][t]{\fontsize{8}{12}\fontseries{m}\mathversion{normal}\fontshape{n}\selectfont \color[rgb]{0,0,0}\setlength{\tabcolsep}{0pt}\begin{tabular}{c}$\test$ [ms]\end{tabular}}%
\psfrag{s10}[][]{\fontsize{10}{15}\fontseries{m}\mathversion{normal}\fontshape{n}\selectfont \color[rgb]{0,0,0}\setlength{\tabcolsep}{0pt}\begin{tabular}{c} \end{tabular}}%
\psfrag{s11}[][]{\fontsize{10}{15}\fontseries{m}\mathversion{normal}\fontshape{n}\selectfont \color[rgb]{0,0,0}\setlength{\tabcolsep}{0pt}\begin{tabular}{c} \end{tabular}}%
\psfrag{s12}[l][l]{\fontsize{8}{12}\fontseries{m}\mathversion{normal}\fontshape{n}\selectfont \color[rgb]{0,0,0}Corollary 1}%
\psfrag{s13}[l][l]{\fontsize{8}{12}\fontseries{m}\mathversion{normal}\fontshape{n}\selectfont \color[rgb]{0,0,0}IM}%
\psfrag{s14}[l][l]{\fontsize{8}{12}\fontseries{m}\mathversion{normal}\fontshape{n}\selectfont \color[rgb]{0,0,0}EM-AC, Thm. 1}%
\psfrag{s15}[l][l]{\fontsize{8}{12}\fontseries{m}\mathversion{normal}\fontshape{n}\selectfont \color[rgb]{0,0,0}EM-OC, Thm. 2}%
\psfrag{s16}[l][l]{\fontsize{8}{12}\fontseries{m}\mathversion{normal}\fontshape{n}\selectfont \color[rgb]{0,0,0}Corollary 1}%
%
\fontsize{8}{12}\fontseries{m}\mathversion{normal}%
\fontshape{n}\selectfont%
%
\psfrag{x01}[t][t]{1}%
\psfrag{x02}[t][t]{2}%
\psfrag{x03}[t][t]{3}%
\psfrag{x04}[t][t]{4}%
\psfrag{x05}[t][t]{5}%
\psfrag{x06}[t][t]{6}%
\psfrag{x07}[t][t]{7}%
\psfrag{x08}[t][t]{8}%
\psfrag{x09}[t][t]{9}%
\psfrag{x10}[t][t]{10}%
%
\psfrag{v01}[r][r]{0.8}%
\psfrag{v02}[r][r]{0.82}%
\psfrag{v03}[r][r]{0.84}%
\psfrag{v04}[r][r]{0.86}%
\psfrag{v05}[r][r]{0.88}%
\psfrag{v06}[r][r]{0.9}%
\psfrag{v07}[r][r]{0.92}%
\psfrag{v08}[r][r]{0.94}%
\psfrag{v09}[r][r]{0.96}%
\psfrag{v10}[r][r]{0.98}%
\psfrag{v11}[r][r]{1}%
%
%

\begin{tikzpicture}[scale=1]
\node[anchor=south west,inner sep=0] (image) at (0,0)
{
\includegraphics[width = \figscale]{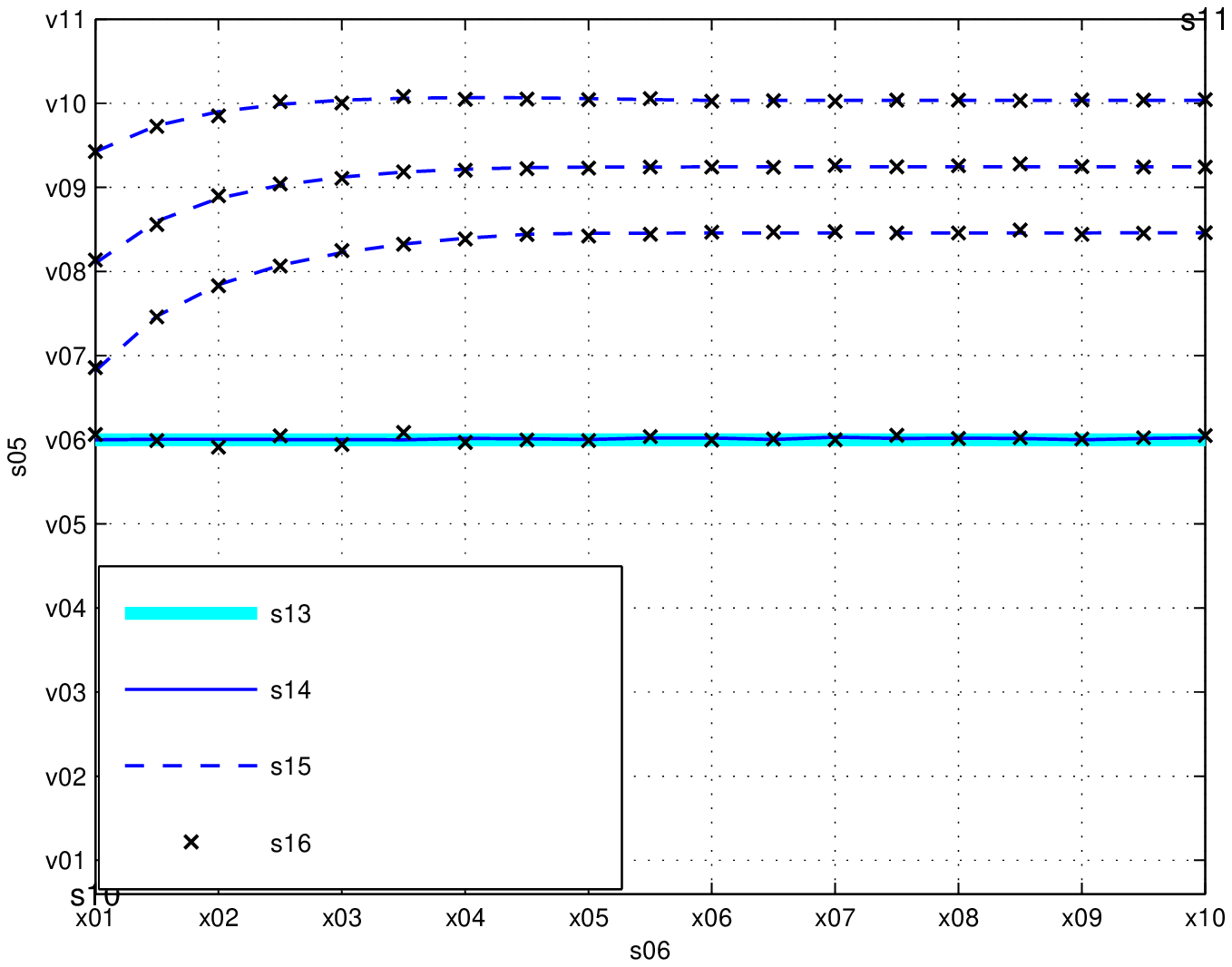} 
};
\begin{scope}[x={(image.south east)},y={(image.north west)}]
\draw[black,->] (0.25,0.68) -- (0.14,0.92);
\node[draw=none,font=\scriptsize] at (0.25,0.62) {$\mpd \in \{0.05,0.10,0.15\}$}; 

\end{scope}
\end{tikzpicture}

\label{fig:pd_test}}
\hfil
\subfloat[]{
%
%
%
\psfrag{s05}[b][b]{\fontsize{8}{12}\fontseries{m}\mathversion{normal}\fontshape{n}\selectfont \color[rgb]{0,0,0}\setlength{\tabcolsep}{0pt}\begin{tabular}{c}$\pfa$\end{tabular}}%
\psfrag{s06}[t][t]{\fontsize{8}{12}\fontseries{m}\mathversion{normal}\fontshape{n}\selectfont \color[rgb]{0,0,0}\setlength{\tabcolsep}{0pt}\begin{tabular}{c}$\test$ = [ms]\end{tabular}}%
\psfrag{s10}[][]{\fontsize{10}{15}\fontseries{m}\mathversion{normal}\fontshape{n}\selectfont \color[rgb]{0,0,0}\setlength{\tabcolsep}{0pt}\begin{tabular}{c} \end{tabular}}%
\psfrag{s11}[][]{\fontsize{10}{15}\fontseries{m}\mathversion{normal}\fontshape{n}\selectfont \color[rgb]{0,0,0}\setlength{\tabcolsep}{0pt}\begin{tabular}{c} \end{tabular}}%
\psfrag{s12}[l][l]{\fontsize{8}{12}\fontseries{m}\mathversion{normal}\fontshape{n}\selectfont \color[rgb]{0,0,0}Corollary 1}%
\psfrag{s13}[l][l]{\fontsize{8}{12}\fontseries{m}\mathversion{normal}\fontshape{n}\selectfont \color[rgb]{0,0,0}IM}%
\psfrag{s14}[l][l]{\fontsize{8}{12}\fontseries{m}\mathversion{normal}\fontshape{n}\selectfont \color[rgb]{0,0,0}EM-AC, Thm. 1}%
\psfrag{s15}[l][l]{\fontsize{8}{12}\fontseries{m}\mathversion{normal}\fontshape{n}\selectfont \color[rgb]{0,0,0}EM-OC, Thm. 2}%
\psfrag{s16}[l][l]{\fontsize{8}{12}\fontseries{m}\mathversion{normal}\fontshape{n}\selectfont \color[rgb]{0,0,0}Corollary 1}%
%
\fontsize{8}{12}\fontseries{m}\mathversion{normal}%
\fontshape{n}\selectfont%
%
\psfrag{x01}[t][t]{1}%
\psfrag{x02}[t][t]{2}%
\psfrag{x03}[t][t]{3}%
\psfrag{x04}[t][t]{4}%
\psfrag{x05}[t][t]{5}%
\psfrag{x06}[t][t]{6}%
\psfrag{x07}[t][t]{7}%
\psfrag{x08}[t][t]{8}%
\psfrag{x09}[t][t]{9}%
\psfrag{x10}[t][t]{10}%
%
\psfrag{v01}[r][r]{$10^{-4}$}%
\psfrag{v02}[r][r]{$10^{-3}$}%
\psfrag{v03}[r][r]{$10^{-2}$}%
\psfrag{v04}[r][r]{$10^{-1}$}%
%
%

\begin{tikzpicture}[scale=1]
\node[anchor=south west,inner sep=0] (image) at (0,0)
{
\includegraphics[width = \figscale]{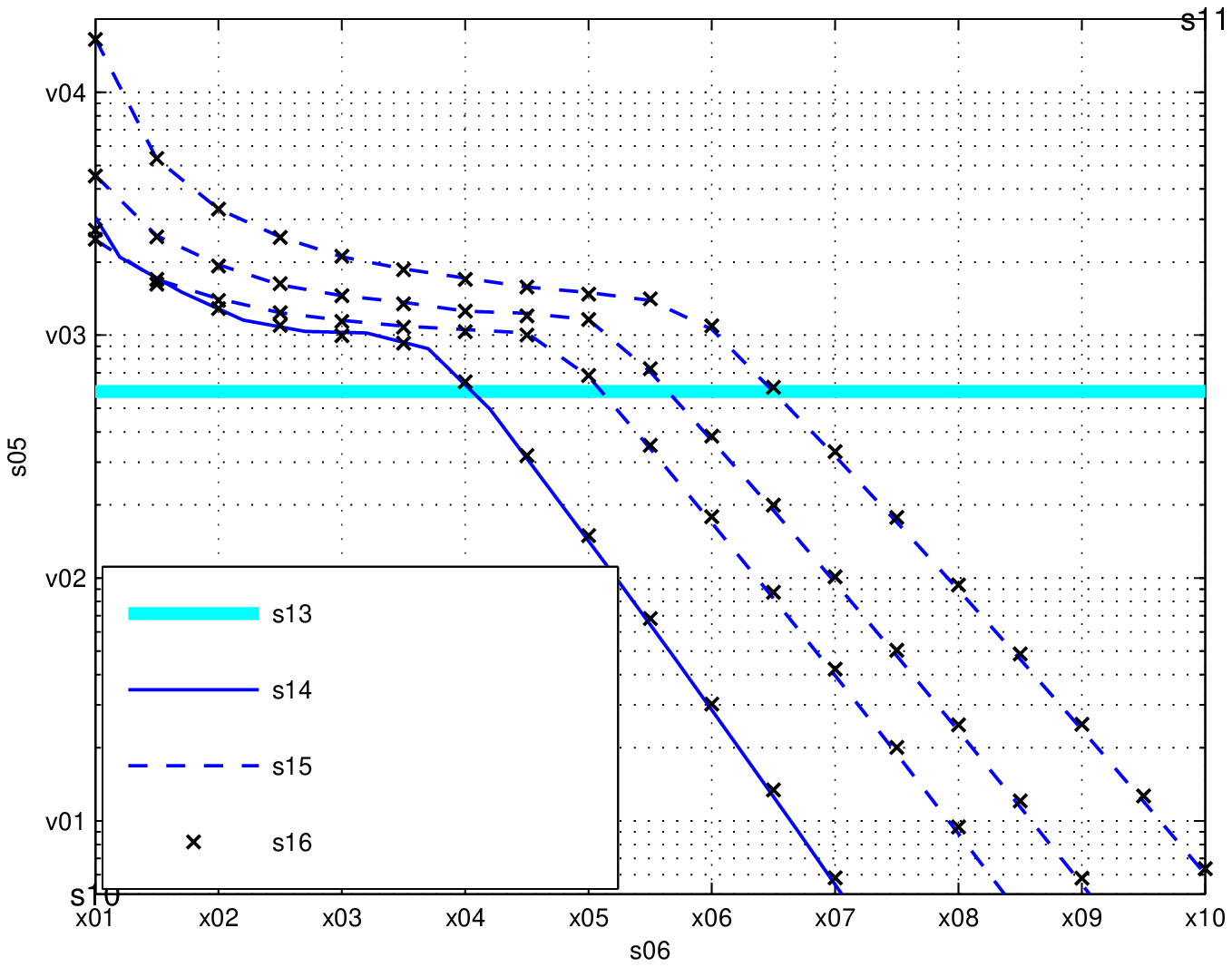} 
};
\begin{scope}[x={(image.south east)},y={(image.north west)}]
\draw[black,->] (0.75,0.47) -- (0.6,0.33);
\node[draw=none, rotate=-50, font=\scriptsize] at (0.78,0.5) {$\mpd \in \{0.05,0.10,0.15\}$}; 

\end{scope}
\end{tikzpicture}

\label{fig:pf_tsen}}
\vspace{0.3cm}
\caption{\tc{Variation of $\e{\pd}{\pd}$ and $\pfa$ versus the $\test$, where the secondary throughput is maximized over the sensing time, $\trs(\test,\ttsen)$. (a) Expected $\pd$ versus $\test$, (b) $\pfa$ versus $\test$.}} 
\label{fig:ROC_test}
\end{figure}

To procure further insights, we investigate the variations of expected $\pd$ and $\pfa$ with the estimation time. From \figurename~\ref{fig:pd_test}, it is observed that the expected $\pd$ corresponding to the outage constraint is strictly above the desired level $\pdd$ for all values of estimation time, however, for lower values of estimation time, this margin reduces. This is based on the fact that lower estimation time shifts the probability mass of $\pd$, to a lower value, \tc{refer to} \figurename~\ref{fig:CDF_pd_test}. 
\tc{According to \figurename~\ref{fig:pf_tsen}, the system notices a considerable improvement in $\pfa$ at small values of $\test$, which saturates for a certain period and falls drastically beyond a certain value. To understand this, it is important to study the dynamics between the estimation and the sensing time. Low $\test$ increases the variations in the detection probability, these variations are compensated by an increase in the suitable sensing time, and vice versa. The performance improves until a maximum ($\ttest$, $\ttsen$) is reached, beyond this, the time resources (allocated in terms of the sensing and the estimation time) contribute more in improving the detector's performance (in terms of $\pfa$ as $\pd$ is already constrained) and less in reducing the variations due to channel estimation.}

\section{Conclusion} \label{sec:conc}
In this paper, we have investigated the performance of cognitive radio as an interweave system from a deployment perspective. It has been argued that the knowledge of the interacting channels is a key aspect that enables the performance characterization of the interweave system in terms of sensing-throughput tradeoff. In this regard, a novel framework that facilitates channel estimation and captures the effect of estimation in the system model has been proposed. As a major outcome of the analysis, it has been justified that the existing model, illustrating an ideal scenario, overestimates the performance of the interweave system, hence, less suitable for deployment. Moreover, it has been clearly stated that the variations induced in the system, specially in the detection probability may severely degrade the performance of the primary system. To overcome this situation, average and outage constraints as primary user constraints have been employed. As a consequence, for the proposed estimation model, novel expressions for sensing-throughput tradeoff based on the mentioned constraints have been established. More importantly, by analyzing the estimation-sensing-throughput tradeoff, the suitable estimation time and the suitable sensing time that maximize the secondary throughput have been determined. In our future work, we plan to extend the proposed analysis for the hybrid cognitive radio system that combines the advantages of interweave and underlay techniques.  

\section*{\tc{Appendix}}
\subsection{Proof of Lemma \ref{lm:lem3}} 
\begin{IEEEproof}
For simplification, we break down the expression $\left( \frac{|\ehs|^2 \ptranst}{\eprcvdsr} \right)$ in (\ref{eq:Cap1}), as $E_1 = \left( \frac{|\ehs|^2 \ptranst}{\npo} \right)$ and $E_2 = \left(\frac{\eprcvdsr}{\npo} \right)$, where $\co = \log_2 \left(1 + \frac{E_1}{E_2} \right)$. The pdf of the expression $E_1$ is determined in (\ref{eq:dsnrs}).

Following the characterization $\eprcvdsr$ in (\ref{eq:ehp2}), the pdf of $E_2$ is determined as 
\begin{align}
\dsnrp &= \frac{\Kp \npo}{\prcvdsr} \frac{1}{ 2^{\frac{\Kp}{2}} \Gamma\left(\frac{\Kp}{2}\right)} \left(x \frac{\Kp \npo}{ \prcvdsr} \right)^{\frac{\Kp}{2} - 1} \times \nonumber  \\ \quad & \exp \left(- x \frac{\Kp \npo}{2 \prcvdsr}  \right) \label{eq:dsnrp}. 
\end{align}
Using the characterizations of pdfs $\dsnrs$ and $\dsnrp$, we apply Mellin transform \cite{NIST} to determine the pdf of $\frac{E_1}{E_2}$ as
\begin{align}
\dsnrsp(x) &= \frac{x^{\as - 1} \Gamma(\as + \ap)}{\Gamma(\as) \Gamma(\ap) \bs^{\as} \bp^{\ap}} \left(\frac{1}{\bp} + \frac{x}{\bs}\right)^{(\as + \ap)}. \label{eq:dsnrsp} 
\end{align} 
Finally, substituting the expression $\frac{E_1}{E_2}$ in $\co$ yields (\ref{eq:den_C1}).
\end{IEEEproof}

\subsection{Proof of Theorems \ref{th:th1} and \ref{th:th2}} 
\begin{IEEEproof} 
In order to solve the constrained optimization problems illustrated in Theorem \ref{th:th1} and Theorem \ref{th:th2}, the following approach is considered. As a first step, an underlying constraint is employed to determine $\mu$ as a function of the $\tsen$ and $\test$. 

For the average constraint, the expression $\e{\pd}{\pd}$ in (\ref{eq:AC}) did not lead to a closed form expression, consequently, no analytical expression of $\thrac$ is obtained. In this context, we procure $\thrac$ for the average constraint numerically from (\ref{eq:AC}). 

Next, we determine $\throc$ based on the outage constraint. This is accomplished by combining the expression of $\fpd$ in (\ref{eq:fpd}) with the outage constraint (\ref{eq:OC}) 
\begin{align}
P(\pd \le \pdd) = \fpd(\pdd) \le \mpd. 
\label{eq:int1}
\end{align}
Rearranging (\ref{eq:int1}) gives
\begin{align}
\throc &\ge \frac{4 \prcvd \Gamma^{-1} \left(1 - \mpd, \frac{\test \fsam}{2}\right) \Gamma^{-1} \left(\pdd, \frac{\tsen \fsam}{2}\right)}{ \test \tsen (\fsam)^2  }. 
\label{eq:throc}
\end{align}
\tc{Clearly, the random variables $\pd(\eprcvd)$, and $\cz(|\ehs|^2)$ and $\co(|\ehs|^2, \eprcvdsr)$ are functions of the independent random variables $\eprcvd$, and $|\ehs|^2$ and $\eprcvdsr$, respectively. In this context, we apply the independence property on $\pd$, $\cz$ and $\co$ to obtain 
\begin{align*}
\e{\pd, \cz, \co}{\cz (1- \pfa) + \co (1 - \pd)} &= \e{\cz}{\cz} (1 - \pfa) + \\ \quad & \e{\co}{\co} \e{\pd}{(1 - \pd)}\end{align*}
in (\ref{eq:thr_AC}) and (\ref{eq:thr_OC}).} Upon replacing the respective thresholds in $\pd$ and $\pfa$ and evaluating the expectation over $\pd$, $\cz$ and $\co$ using the distribution functions characterized in Lemma \ref{lm:lem1}, Lemma \ref{lm:lem2} and Lemma \ref{lm:lem3}, we determine the expected throughput as a function of sensing and estimation time. 
\end{IEEEproof}
}

\bibliographystyle{IEEEtran}
\bibliography{IEEEabrv,refs}

\end{document}